\documentclass[aps,prx,reprint,amsmath,amssymb,superscriptaddress, nofootinbib, 10pt]{revtex4-2}
\usepackage[english]{babel}
\usepackage{amsmath,amssymb,graphicx,color,comment}
\usepackage[pdfusetitle,bookmarks=true,colorlinks,citecolor=blue,urlcolor=blue, linkcolor=blue]{hyperref}

\usepackage{braket}

\usepackage{subfigure}

\usepackage{extarrows}
\usepackage{chemformula}
\usepackage{float}
\usepackage{mathtools}
\usepackage{diagbox}

\usepackage{adjustbox}
\usepackage{dcolumn}  
\usepackage{bm}       
\usepackage{amsfonts}  
\usepackage{amsmath}
\usepackage{lineno}

\newcommand{\pwisein}{\left\{ \begin{array}{ll}}
	\newcommand{\pwiseout}{\end{array}\right.}

\newcommand{\eq}[1]{Eq.~(\ref{#1})}
\newcommand{\eqq}[2]{Eqs.~(\ref{#1}) and (\ref{#2})}
\newcommand{\eqqs}[2]{Eqs.~(\ref{#1}) and (\ref{#2})}

\newcommand{\fig}[1]{Fig.\thinspace{}\ref{#1}}

\newcommand{\fc}[1]{{#1})}
\newcommand{\figc}[2]{Fig.\thinspace{}\ref{#1}\thinspace{}\fc{#2}}
\newcommand{\figcc}[3]{Fig.\thinspace{}\ref{#1}\thinspace{}\fc{#2} and \fc{#3}}

\newcommand{\supp}{Supplementary Material}
\newcommand{\meth}{Methods}

\newcommand{\Secc}[1]{Section \ref{#1}}
\newcommand{\Tab}[1]{Tab. \ref{#1}}

\usepackage{times}

\newcommand{\titleinfo}{Theory of Nonlinear Spectroscopy of Quantum Magnets}

\begin{document}

\title{\titleinfo}

\author{Anubhav Srivastava}
\affiliation{Technical University of Munich, TUM School of Natural Sciences, Physics Department, 85748 Garching, Germany}
\affiliation{Munich Center for Quantum Science and Technology (MCQST), Schellingstr. 4, 80799 M{\"u}nchen, Germany}
\affiliation{ Indian Institute of Science, Bangalore 560012, India}
\author{Stefan Birnkammer}
\affiliation{Technical University of Munich, TUM School of Natural Sciences, Physics Department, 85748 Garching, Germany}
\affiliation{Munich Center for Quantum Science and Technology (MCQST), Schellingstr. 4, 80799 M{\"u}nchen, Germany}
\author{GiBaik Sim}
\affiliation{Department of Physics, Hanyang University, Seoul 04763, Republic of Korea}
\author{Michael Knap}
\affiliation{Technical University of Munich, TUM School of Natural Sciences, Physics Department, 85748 Garching, Germany}
\affiliation{Munich Center for Quantum Science and Technology (MCQST), Schellingstr. 4, 80799 M{\"u}nchen, Germany}
\author{Johannes Knolle}
\affiliation{Technical University of Munich, TUM School of Natural Sciences, Physics Department, 85748 Garching, Germany}
\affiliation{Munich Center for Quantum Science and Technology (MCQST), Schellingstr. 4, 80799 M{\"u}nchen, Germany}
\affiliation{Blackett Laboratory, Imperial College London, London SW7 2AZ, United Kingdom}

\begin{abstract}
Two-dimensional coherent spectroscopy (2DCS) is an established method for characterizing molecules and has been proposed in the THz regime as a new tool for probing exotic excitations of quantum magnets; however, the precise nature of the coupling between pump field and spin degrees of freedom has remained unclear. Here, we develop a general response theory of 2DCS and show how magneto-electric as well as polarization couplings contribute to 2DCS in addition to the typically assumed magnetization. We propose experimental protocols to distill individual contributions, for instance from exchange-striction or spin current mechanism, when the electric field couples to terms quadratic in spin operators. We provide example calculations for the paradigmatic twisted Kitaev chain material $\ch{CoNb2O6}$ and highlight the crucial role of contributions from cross-coupling between polarization and magnetic nonlinear susceptibilities. Our work paves the way for systematic studies of light-matter couplings in quantum magnets and for establishing 2DCS as a versatile tool for probing fractional excitations of exotic magnetic quantum phases. 
\end{abstract}

\date{\today}
\maketitle
\textbf{Introduction:} Spectroscopic measurements are one of the main experimental tools for understanding the microscopic constituents of different forms of matter~\cite{devereaux2007inelastic}. 2D coherent spectroscopy (2DCS), a two-pump probe nonlinear method, not only probes intrinsic excitations but also their interactions~\cite{mukamel1995principles}. 2DCS is well established for studying molecules and proteins~\cite{hamm2011concepts,pavia2015introduction} but recent advances in the generation of THz pulses have allowed the study of excitation energies relevant for quantum materials~\cite{Woerner2013}. 2DCS is particularly promising for probing quantum magnets with exotic fractionalized excitations~\cite{Wan2019,choi2020theory,Hart2023,fava2023divergent,Sim2023a,Sim2023b,mcginley2024signatures,Watanabe2024,zhang2024disentangling,potts2024signatures}, the latter being hard to detect in linear spectroscopy as there they lead to broad continua which could also arise from mundane effects like disorder~\cite{zhu2017disorder} or thermal smearing~\cite{franke2022thermal}. The recent observations of magnon excitations~\cite{lu2017coherent} and their interactions~\cite{zhang2024terahertz} in insulating ordered magnets via 2DCS confirm the power of the method, but so far, the microscopic light-matter couplings involved in 2DCS remain unclear to a large extent. \\
Thus far numerous theoretical 2DCS studies for quantum magnets~\cite{Wan2019,choi2020theory,Hart2023,fava2023divergent,Sim2023a,Sim2023b,mcginley2024signatures,Watanabe2024,zhang2024disentangling,potts2024signatures} have solely focused on the nonlinear \emph{magnetization response}. However, when considering a free electron coupled to an electromagnetic wave,  both the electric as well as the magnetic components couple. Assuming the free electron to be delocalized on the scale of the Bohr radius, the electric coupling even dominates over the magnetic coupling by a factor of the inverse fine structure constant, $1/\alpha \approx 137$. Of course, the situation is more complex for localized electrons in Mott insulators with only spin excitations at low energy. It is nonetheless pertinent to investigate its contribution to 2DCS, because strong intrinsic electrical polarization of spin origin have also been established in quantum magnets~\cite{Curie1894,tokura2014multiferroics,bulaevskii2008electronic,bordacs2012chirality,katsura2005spin,Nandi2019}. \\
Motivated by these considerations, in this work, we provide a microscopic theory of 2DCS of quantum magnets and show that the coupling of the THz field to the intrinsic electric polarization gives an important yet overlooked contribution to the nonlinear response. We discuss the distinct signatures of the electric polarization in 2DCS and how to separate its contribution in experiments. We provide example calculations for a 1D soluble twisted Kitaev model, motivated by the material~$\ch{CoNb2O6}$~\cite{Kjaell_2011,Fava2020,Morris2021}.
\begin{figure}[t!]
\includegraphics[width=1\columnwidth]{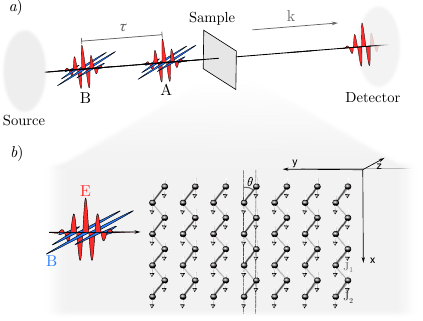}
\caption{\textbf{Setup for 2DCS of quantum magnets.} (a) For measuring the nonlinear response a sample is exposed to two consecutive 
collinear pump pulses $A$ and $B$ from a source with propagation vector $k$. Interactions with electric $\vec{E}$ and magnetic $\vec{B}$ field components of the pulse induce a non-vanishing polarization as well as magnetization in the sample. The resulting electric field emitted from the sample can be measured experimentally in a detector via electro-optical sampling (EOS). (b) We analyze the response of a dimerized twisted Kitaev chain as a candidate quantum magnet realized by the compound $\ch{CoNb2O6}$. To exploit the solvable nature of the model, the THz pulse geometry is chosen such that the electric field is aligned with the $\hat{x}$-axis while the magnetic field is aligned with the $\hat{z}$-axis~\cite{Sim2023a}.}
\label{fig1} 
\end{figure}

\textbf{Setup of 2DCS:} 
First, we introduce the conventionally used protocol for 2DCS~\cite{Wan2019,pavia2015introduction} illustrated in \figc{fig1}{a}, which consists of two collinear consecutive THz pulses, pumps $A$ and $B$,  on the sample. For simplicity, we assume the pulses $A$ and $B$ to be well approximated by $\delta$-functions at times $t=0$ and $t=\tau$ for a given orientation
\begin{align}
    \vec{E}(t) = E_{0}\delta(t)\hat{x} + E_{\tau}\delta(t - \tau)\hat{x}  \label{eq:PulseSequenceE}\\
    \vec{B}(t) = B_{0}\delta(t)\hat{z} + B_{\tau}\delta(t - \tau)\hat{z}. 
\label{eq:PulseSequence}
\end{align}
In general the amplitudes of both pump pulses can be different but we will consider the same intensities for simplicity (i.e. $E_{0}=E_{\tau}, B_{0}=B_{\tau}$). The electric field (amplitude and phase) of the emitted THz pulse is measured at a later time $t+\tau$ by electro-optic sampling~\cite{Woerner2013,lu2017coherent}.\\
 Crucially, in the far field, we argue that the measured electric field consists of contributions not only from the magnetization $M(t)$ but also the polarization $P(t)$~\cite{Jackson1999}, i.e.
\begin{equation}
    E(t) = i\bigl[a_{E}  P(t) + b_{E} M(t)\bigl].
    \label{eq:dipolerad}
\end{equation}
The weighting factor $a_{E}, b_{E}$ thereby take into account sample-specific parameters and geometric properties of the setup. The resulting magnetic field of the signal has an analogous expression; however, it is typically not measured in EOS experiments. Investigating the nonlinear response of the system requires to repeat the experiment two more times with one of the two pump pulses $A$ and $B$ only. Subtracting the two pump-probe measurements from the measurement with both pulses then yields the nonlinear 2DCS response~\cite{Wan2019}.\\
Next, we relate changes in polarization (and magnetization) to applied pumps with susceptibilities $\chi^{P (n)}$ (and $\chi^{M (n)}$) of different order $n$~\cite{mukamel1995principles} using time-dependent perturbation theory.  For most cases, the measured signal is dominated by second-order responses in polarization and magnetization
\begin{widetext}
\vspace{-\baselineskip}
\begin{align}
P^{\text{NL}}(t+\tau)  &=  \chi^{P (2)}_{PP}(t, t + \tau)  E_{\tau}E_{0} +  \chi^{P (2)}_{MP}(t, t + \tau)B_{\tau}E_{0} + 
\chi^{P (2)}_{PM}(t, t + \tau) E_{\tau}B_{0} +\chi^{P (2)}_{MM}(t, t + \tau) B_{\tau}B_{0}  + \dots \label{eq:P_NL} \\
M^{\text{NL}}(t+\tau) &=   \chi^{M (2)}_{PP}(t, t + \tau) E_{\tau} E_{0} +  \chi^{M (2)}_{PM}(t, t + \tau)E_{\tau}B_{0} +
\chi^{M (2)}_{MP}(t, t + \tau) B_{\tau}E_{0} +\chi^{M (2)}_{MM}(t, t + \tau) B_{\tau}B_{0} + \dots  . \label{eq:M_NL} 
\end{align}
\end{widetext}
At a given order $n$ we obtain $2^{n}$ susceptibilities taking into account interactions with the electric and magnetic field components of the pumps, respectively. The dots represent higher-order non-linear response. 
In most non-linear spectroscopy experiments, $\chi^{(3)}$ is the generic non-linear response, but for systems with reduced symmetry the dominant non-linear contribution can arise from $\chi^{(2)}$~\cite{lu2017coherent}. In the following, we first focus on the case with a dominating $\chi^{(2)}$ response when the system breaks local symmetries, i.e. for our example of the twisted Kitaev chain, it is the absence of the glide symmetry as discussed below. We later analyze the case of dominating $\chi^{(3)}$ relevant for systems with glide symmetry.\\
Most previous theoretical works on quantum magnets have focused on direct interactions between the spin degrees of freedom and the magnetic field of the pulse, thus, probing only the magnetic responses like $\chi^{M (2)}_{MM}$~\cite{Wan2019,choi2020theory,Hart2023,fava2023divergent,Sim2023a,Sim2023b,mcginley2024signatures,Watanabe2024,zhang2024disentangling,potts2024signatures,Qiang2024,David2025}. A step further was recently taken in Ref.~\cite{Brenig2024} studying 
$\chi^{P (2)}_{PP}$ for the Kitaev QSL.  Here, our goal is to provide the general response theory of quantum magnets elucidating the role of all additional contributions to the measurement signal arising from finite polarizations~\cite{bulaevskii2008electronic,katsura2005spin,Bolens2018,You2014}. Polarization of spin systems arises from different microscopic mechanisms~\cite{tokura2014multiferroics} and we will concentrate on (i) exchange-striction~\cite{bulaevskii2008electronic} or (ii) spin-current mechanism~\cite{katsura2005spin}. 
While the first term refers to polarization which can be externally induced via a DC electric field, the second contribution is a result of spin-orbit coupling. The magnitude of exchange-striction effects is thereby related to the amount of inversion symmetry breaking in the material, which, if present, usually dominates spin-current effects\cite{Dong2019,tokura2014multiferroics,katsura2005spin}. The form of the polarization operator in terms of spin follows from symmetry constraints, e.g. being odd under inversion and even under time reversal, but the precise form and coupling strength is material specific~\cite{Miyahara2016}. Below, we discuss the consequences of symmetries by means of a specific example.

\textbf{Isolating polarization contributions:} 
In general, a 2DCS experiment following the protocol of \figc{fig1}{a} measures the sum of all different contributions, including the conventionally considered magnetization susceptibility $\chi^{M (2)}_{MM}$, the polarization $\chi^{M (2)}_{PP}$, and also cross-correlations $\chi^{M (2)}_{PM}$ and $\chi^{M (2)}_{MP}$ (plus similar contributions from $\chi^{P (2)}_{...}$ see \eq{eq:P_NL}). 
Exploiting symmetry properties we can derive as a key result an experimental protocol to isolate different susceptibilities contained in \eqqs{eq:P_NL}{eq:M_NL} at second order, which allows us to provide a direct experimental test to quantify the magnitude of cross-correlations and polarization coupling.\\
In order to isolate specific contributions to the response one can superimpose the measurement outcomes of several experimental runs with different parameters. As an example, we consider inverting the electric field component of the probe pulses, which can be achieved by inverting the direction of pulse propagation or by rotating the sample see \meth, resulting in modified polarization $P^{\text{NL}}(t\vert -\vec{E}, \vec{B})$ and magnetization $M^{\text{NL}}(t\vert -\vec{E}, \vec{B})$ compared to previously measured quantities $P^{\text{NL}}(t\vert \vec{E}, \vec{B})$ and $M^{\text{NL}}(t\vert \vec{E}, \vec{B})$. Note, due to the sign change all components linear in the inverted field component pick up a global sign and the combination of measurements provides access to
\begin{align}
    P^{\text{NL}}_{\parallel}(t) &\equiv \bigl(P^{\text{NL}}(t\vert \vec{E}, \vec{B}) + P^{\text{NL}}(t\vert -\vec{E}, \vec{B})\bigr)/2  \nonumber\\ 
    &= \chi^{P (2)}_{PP}(t-\tau, t)E_{\tau}E_{0}  + \chi^{P (2)}_{MM}(t -\tau, t)B_{\tau}B_{0}  + \dots \label{eq:P_parallel}
\end{align}
\begin{align}
    P^{\text{NL}}_{\times}(t) &\equiv \bigl(P^{\text{NL}}(t\vert \vec{E}, \vec{B}) - P^{\text{NL}}(t\vert -\vec{E}, \vec{B})\bigr)/2 \nonumber\\
    &= \chi^{P (2)}_{MP}(t-\tau, t ) B_{\tau}E_{0}  + \chi^{P (2)}_{PM}(t -\tau, t)E_{\tau}B_{0}  + \dots .\label{eq:P_cross}
\end{align}
Equivalent relations can also be formulated for the magnetization $M^{\text{NL}}_{\parallel / \times}$, see \supp. The absence of polarization couplings would lead to a vanishing $P^{\text{NL}}_{\times}$, although $P^{\text{NL}}_{\parallel}$ would be finite due to the purely magnetic contribution.
However, for an experimental verification we need to study the combination of measurement signals. For this, it is required to first analyze the transformation properties of $a_E$ and $b_E$ under inversion of the pulse propagation geometry. Overall, the transformation respects the parity symmetry of electrodynamics. Combining this with the vector (pseudo-vector) like properties of $\vec{E}, \vec{P}$ ($\vec{M}$) being odd (even) under parity transformations requires $b_{E}$ to flip its sign after inverting the electric field component, while $a_{E}$ remains unaltered, see Eq.~\eqref{eq:dipolerad}. Superimposing the different measurement outcomes with opposite electric field directions, $E(t\vert \vec{E}, \vec{B})$ and $E(t\vert -\vec{E}, \vec{B}) = i\bigl[a_{E}  P(t\vert -\vec{E}, \vec{B}) - b_{E} M(t \vert -\vec{E}, \vec{B})\bigl]$ yields

\begin{align}
    E^{\text{NL}}_{\text{sym}}(t) &= \bigl(E^{\mathrm{NL}}(t\vert \vec{E}, \vec{B}) + E^{\mathrm{NL}}(t\vert -\vec{E}, \vec{B})\bigr)/2 \nonumber \\
    &=  i\bigl[ a_{E} P^{\mathrm{NL}}_{\parallel} (t) +  b_{E} M^{\mathrm{NL}}_{\times} (t)\bigr ] \label{eq:E_NL_sym}\\
    E^{\mathrm{NL}}_{\text{asym}}(t) &= \bigl(E^{\mathrm{NL}}(t\vert \vec{E}, \vec{B}) - E^{\mathrm{NL}}(t\vert -\vec{E}, \vec{B}) \bigr)/2\nonumber \\
    &=  i\bigl[ a_{E} P^{\mathrm{NL}}_{\times} (t) +  b_{E} M^{\mathrm{NL}}_{\parallel} (t)\bigr].
    \label{eq:E_NL_asym}
\end{align} 
Crucially, as the symmetric response $E^{\mathrm{NL}}_{\mathrm{sym}}(t)$ only contains terms involving a polarization component, thus, a non-vanishing response serves as a direct indicator for finite polarization coupling in a material, for details see \meth .

\begin{figure*}[t]
\includegraphics[width=\textwidth]{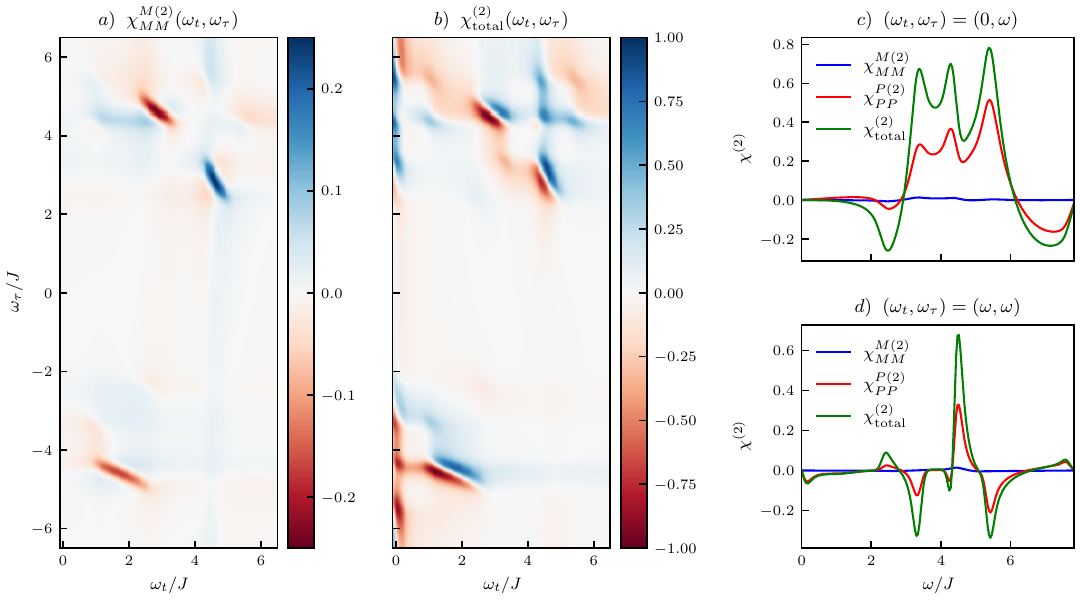}
  \caption{\textbf{Second order response of a dimerized TKSC.} We compare (a) the pure magnetic response $\chi^{M (2)}_{MM}$ against (b) the total response $\chi^{(2)}_{\mathrm{total}}= \chi^{M (2)}_{MM} + \chi^{M (2)}_{PM} + \chi^{M (2)}_{MP} + \chi^{M (2)}_{PP} + \chi^{P (2)}_{MM} + \chi^{P (2)}_{PM} + \chi^{P (2)}_{MP} + \chi^{P (2)}_{PP}$ including interactions with the electric field via exchange-striction (ES) at second order. Note the different scales of the color bars in (a) and (b). (c)-(d) High-symmetry cuts through the two-dimensional frequency space $(\omega_{t}, \omega_{\tau})$ indicate additional signatures in the measurement response from electric and cross coupling terms, absent in the pure magnetic response. (c) Along the $\omega_{\tau}$-axis ($\omega_{t}\!=\!0$) we can discriminate a vanishing response for magnetic only coupling against nontrivial spectra of the polarization response $\chi^{P (2)}_{PP}$ (red) and total response $\chi^{(2)}_{\mathrm{total}}$. (d) Qualitatively similar results are found for diagonal cuts of the frequency plane ($\omega_{t}=\omega_{\tau}$). While characteristic signatures in the spectrum are present in the pure polarization response $\chi^{P (2)}_{PP}$ and even amplified in the total response $\chi^{ (2)}_{\mathrm{total}}$, the magnetization response $\chi^{M (2)}_{MM}$ is featureless on comparable scales. For comparison of polarization and magnetization contributions we consider equal geometric weighting factors $a_{E}=b_{E}$. Moreover, all results are derived for finite dimerization in the couplings of $J_{2}=1.5 J_{1}$.}
\label{fig2}
\end{figure*}
\textbf{Example response for Kitaev chain material:} After our general discussion we next provide an illustrative example. We analyze the response for a paradigmatic quasi one-dimensional quantum magnet, i.e. the twisted Kitaev spin chain (TKSC) as a basic minimal model of the compound $\ch{CoNb2O6}$ reduced to a single zig-zag chain~\cite{Fava2020,Morris2021}, see \figc{fig1}{b}. While the material realization of the compound has a non-aligned local and crystal axes~\cite{Morris2021,Sim2023a}, our minimal model neglects this subtlety but is meant to serve as a qualitative soluble example to illustrate the distinct coupling contributions.  In our model interactions are governed by 
\begin{equation}
    H_{\mathrm{TK}} =-\sum_{i=1}^{L/2} \Big[ J_1\tilde{\sigma}_{2i-1}(\theta)\tilde{\sigma}_{2i}(\theta)+J_2\tilde{\sigma}_{2i}(-\theta)\tilde{\sigma}_{2i+1}(-\theta)\Big] .
\label{eq:TKSC}
\end{equation}
Here we introduce a rotated basis for the spin-$\frac{1}{2}$ variables to simplify the notation of interactions between neighboring spins, i.e.
$\tilde{\sigma}_{i}(\theta)=\cos(\theta)\sigma_{i}^{x}+\sin(\theta)\sigma_{i}^{y}$. $J_1, J_2$ denote positive ferromagnetic exchange couplings, which in the presence of a finite dimerization can be of different strength, for example as a result of an externally applied electric field~\cite{Kanega2021,Brenig2023,Brenig2024}. $H_{\mathrm{TK}}$ is exactly solvable via a Jordan-Wigner transformation in a basis of spinless fermions. The ground state phase diagram interpolates between two canted Ising ferromagnetic phases orientated along the $\hat{x}$- or $\hat{y}$-axes 
via tuning the zig-zag angle $\theta$~\cite{Sim2023a,Sim2023b,Birnkammer2024}.
\begin{figure*}[t!] 
\includegraphics[width=\textwidth]{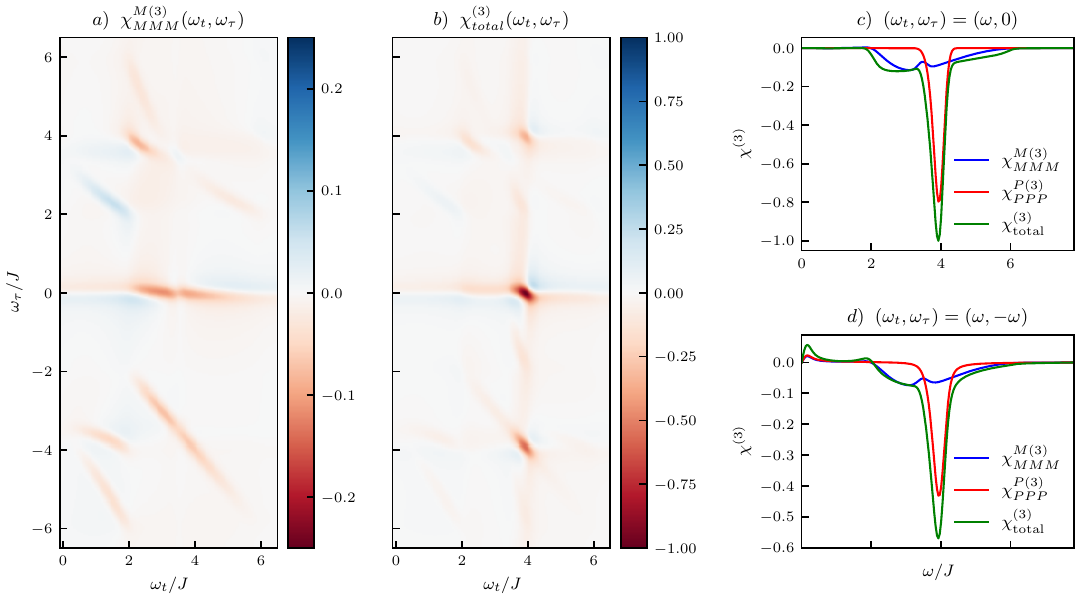}
  \caption{\textbf{Leading third order non-linear response of undimerized TKSC.} (a)-(b) Comparison between the leading order contribution to the non-linear response of an undimerized ($J_{1}=J_{2}$) TKSC. We contrast the response spectrum expected from (a) pure magnetization coupling $\chi^{M (3)}_{MMM}$ (see \meth~for definition) against (b) the total response including all possible interactions with polarization $P_{\mathrm{SC}}$ arising from spin-current effects. The latter adds sharp response features to the otherwise diffuse signatures. Note the different scales of the color bars in (a) and (b). (c)-(d) A quantitative comparison along the (c) horizontal ($\omega_{\tau}=0$) and (d) anti-diagonal ($\omega_{t}=-\omega_{\tau}$) reveal sharp peaks in the pure polarization response $\chi^{P (3)}_{PPP}$ (see \meth~for definition) (red solid lines) that are also reflected in the total signal (green solid line). For comparison of polarization and magnetization contributions we consider equal weighting factors $a_{E}=b_{E}$.}
\label{fig3}
\end{figure*}

\noindent To make use of the exact solubility of the system we chose the geometry of the spectroscopic setup such that the linearly polarized THz pulse propagates against the $\hat{y}$-direction, i.e. $\vec{k}\propto -\hat{y}$. We emphasize that our considerations above do not depend on these specific choices of the geometry. The electric field points along $\hat{x}$ and magnetic field components are in $\hat{z}$-direction, see \figc{fig1}{b} for an illustration. For this setup, the magnetic field couples directly via  $M^{z}B^{z}$. Coupling the quantum spin system to the electric field via $E^{x} P^{x}$ requires a non-zero polarization of the system.\\
Next, we identify microscopic polarizations in terms of spin operators. First, finite polarization of the sample can result from exchange striction (ES), if the system breaks inversion symmetry about the bond center. 
Second, polarization can emerge spontaneously in the system through the spin current (SC) mechanism which is most pronounced in materials with strong spin-orbit coupling~\cite{Dong2019,katsura2005spin,Menchyshyn2015,Bolens2018}, thus also suitable for $\ch{CoNb2O6}$. For our TKSC Hamiltonian~\eqref{eq:TKSC} both contributions take the following form
\begin{align}
    P^x_{\mathrm{ES}} &\propto \sum_{i=1}^{L/2}\bigl( \tilde{\sigma}_{2i-1}(\theta)\tilde{\sigma}_{2i}(\theta)-\tilde{\sigma}_{2i}(-\theta)\tilde{\sigma}_{2i+1}(-\theta)\bigr) 
    \label{eq:PES} \\
    P^x_{\mathrm{SC}} &\propto \sum_{i=1}^{L}(-1)^{i} \sin(\theta)(\sigma_i^{y}\sigma_{i+1}^{x}-\sigma_i^{x}\sigma_{i+1}^{y}).
    \label{eq:PSC}
\end{align}
Here we omitted material specific coupling constants related to the amount of symmetry breaking due to dimerization as well as the magnitude of spin orbit coupling in the material. It is, however, worth noting that ES typically dominates over the polarization resulting from SC effects in systems of broken inversion symmetry~\cite{Dong2019,tokura2014multiferroics}.
The functional form of the coupling enables us to compute the response of the system within the formalism of nested commutators, i.e. for the magnetic susceptibilities 
\begin{equation}
    \chi^{M (2)}_{MM} (t, t + \tau) \equiv - \theta(t) \theta(\tau) \langle [[M^{z}(t + \tau), M^{z}(\tau)], M^{z}(0)] \rangle
    \label{eq:CommutatorStructure}
\end{equation}
and similar for the polarization and cross-couplings all contributing to the polarization $P^{\text{NL}}(t)$ and magnetization $M^{\text{NL}}(t)$ response of \eqqs{eq:P_NL}{eq:M_NL}, for details see \meth .

\textbf{Second order response}: First, we concentrate on the second order response with polarization originating from ES. A nonzero $P_{\mathrm{ES}}$ will arise if the inversion symmetry about the bond center is broken. To get a dominating $\chi^{(2)}$ we need to analyze the symmetries of our model. For $J_1=J_2$, the system has glide symmetries, $G_{y}=T_{c}e^{i\pi/2\sum_{i}^{L}\sigma_{i}^{y}}$ and $G_{x}=T_{c}e^{i\pi/2\sum_{i}^{L}\sigma_{i}^{x}}$ with $T_c$ the translation operator by half a unit cell~\cite{Fava2020}. Since the ground state preserves $G_x$, it acts trivially on it. The coupling terms $P_\text{ES}$ and $M^{z}$ are odd under $G_x$, implying that the even order susceptibilities vanish~\cite{Sim2023a} (see \supp).  To obtain a finite second-order response, we assume that the system has a finite dimerization $(J_{1}/J_{2}\neq 1)$, which explicitly breaks the glide symmetry of the lattice. The dimerization could for example result from a finite DC electric field applied to the system~\cite{Brenig2023,Brenig2024,Kanega2021} or glide symmetry could be intrinsically broken in the material~\cite{You2014,You2016}. \\
In \fig{fig2} we show results for $J_{2}=1.5 J_{1}$  comparing the measurement signals for a pure magnetic 2DCS response as calculated in numerous previous works $\chi^{M (2)}_{MM}$ (panel a), against the total response including also couplings to the electric field $\chi^{(2)}_{\mathrm{total}}$ obtained  from Eq.~\eqref{eq:dipolerad} via exchange striction $P_{\mathrm{ES}}$ at second order  (panel b). To allow for a comparison of both results we assume equal weights for polarization and magnetization, i.e. $a_{E}=b_{E}=1$ and set all unknown coupling constants to order unity. Interestingly, we observe clear qualitative differences between both responses, for example the  additional features for negative frequencies $\omega_{\tau}$ can be attributed to cross coupling terms. A more quantitative comparison is depicted in \figcc{fig2}{c}{d} with high symmetry cuts in frequency space. Both the response of vanishing frequency ($\omega_{t}=0$) as well as the diagonal response ($\omega_{\tau}=\omega_{t}$) show no characteristic features in the conventionally considered magnetic channel $\chi^{M (2)}_{MM}$. In contrast, contributions arising from exchange striction (red lines) in \figcc{fig2}{c}{d} show three distinct peaks. From the exact solution these can be understood as distinct dynamical transitions between fermionic excitation bands~\cite{Sim2023b}. The diagonal peaks, oscillating in $t+\tau$, correspond to non-rephasing signals. The peaks along the $\omega_\tau$ axis come from signals oscillatory only in $\tau$. Viewing $\omega_{\tau}$  as a pumping frequency and $\omega_{t}$ as detection frequency, the signals along the vertical axis represent terahertz rectification signals (R.F.)~\cite{Wan2019,Watanabe2024,lu2017coherent}. The oscillations can take place with frequencies of the fermionic bands of the two-site unit cell. The different signals can then be understood as combinations of spinon pair excitations~\cite{Sim2023b}.

\textbf{Third order response}: Next, we explore potential signatures of SC polarizations in 2DCS. We assume that there is no inversion symmetry breaking such that $P^x_{\mathrm{ES}}$ vanishes and the response results from $P^x_{\mathrm{SC}}$. As discussed above, for unbroken glide symmetry, $J_1=J_2$, the even-order susceptibilities vanish because the symmetry $G_x$ and inversion about bond center act trivially on the ground state of the system, while the coupling terms $M^{z}$ and $P_\text{SC}$ are odd under $G_x$ and bond inversion respectively. Therefore, the leading contribution to the non-linear response arise from third-order susceptibilities. Analogously to the structure of the second-order expression, the third-order response for polarization and magnetization contain eight different contributions each, for details see \meth. Results  for the non-linear response can again be obtained using the solubility of the model and are shown in \fig{fig3}.
We emphasize that while the pure magnetic coupling component $\chi^{M (3)}_{MMM}$ results predominantly in broad features of the response, see \figc{fig3}{a}, coupling to the electric pump field via polarization terms $P_{\mathrm{SC}}$ adds sharp peaks to the spectrum, see \figc{fig3}{b}. Quantitative comparisons are best visualized by horizontal (panel c) and anti-diagonal (panel d) line cuts. Both show sharp resonance contributions from the pure electric coupling $\chi^{P (3)}_{PPP}$, which is also clearly reflected in the total signal $\chi^{(3)}_{\mathrm{total}}$ including also the cross-coupling contributions. The signal along the horizontal $\omega_\tau=0$ axis arises from terms which oscillate only in $t$. If we regard $\omega_\tau$ and $\omega_t$ as pumping and detecting frequency, respectively, this can be interpreted as a pump-probe signal~\cite{Wan2019,lu2017coherent}. The anti-diagonal, so-called spinon echo, signals correspond to perfectly rephasing signals~\cite{Kurnit1964,lu2017coherent,Wan2019}. The signals are oscillatory in $t-\tau$ so that the phase acquired in time $t$ is perfectly canceled by $\tau$. The sharp anti-diagonal peak in coupling to the electric pump field via polarization terms arises because of a peak in the two spinon density of states (DoS). The signals from pure magnetic coupling oscillate with frequencies of twice individual spinon bands because two spinons of the same band are excited. As a result, we find a broad spectrum along the anti-diagonal linecut.

\textbf{Discussion \& Outlook:}
We have shown that 2DCS of quantum magnets not only probes the non-linear magnetization response but also those of the polarization as well as cross-coupling contributions. It will be important to obtain quantitative estimates of the microscopic coupling strengths, for example via ab-initio calculations. 
On the one hand, the hitherto overlooked polarization terms further complicate the interpretation of experimental signatures but we have provided protocols that allow for separating different contributions. In light of these new contributing susceptibilities it would be important to revisit previous theoretical calculations~\cite{Wan2019,choi2020theory,Hart2023,fava2023divergent,Sim2023a,Sim2023b,mcginley2024signatures,Watanabe2024,zhang2024disentangling,potts2024signatures,Qiang2024,David2025}, for example whether signatures of non-trivial braiding statistics can still appear in the total 2DCS signal~\cite{mcginley2024signatures}. On the other hand, the direct coupling of 2DCS to the polarization opens new possibilities for probing quantum magnets. 
Concretely, a non-zero antisymmetric contribution with inverted electric field directions is a direct signature for the presence of magneto-electric couplings. Moreover, in order to obtain large 2DCS signals in experiment one should explore materials with large magneto-electric couplings like in multiferroic hexagonal manganites~\cite{lee2008giant} or recent van der Waals magnets~\cite{zhang2025magnetoelectriceffectvander}.\\
2DCS has been for a long time a powerful method for understanding complex molecules and we hope that our work helps to establish it as a versatile tool for elucidating exotic excitations of quantum magnets. 

\section*{Methods}

\subsection{Susceptibilities from Nested Commutators}

To define the response of the quantum magnet to the external pump pulses we consider a time-dependent perturbation of the system Hamiltonian $H$  
by the electromagnetic field
\begin{equation}
    H(t) = H + \sum_{i} a_{i}(t) X_{i}.
    \label{eq:CouplingH}
\end{equation}
Here $a_{i}(t) X_{i}$ includes polarization coupling to the electric field, i.e. $\vec{E}(t) \cdot \vec{P}$, and magnetization coupling to magnetic field components, i.e. $\vec{B}(t) \cdot \vec{M}$. The response of the system to the drive can be quantified by the evolution of a set of characteristic observables $\{O_{\ell}\}_{\ell}$. From response theory~\cite{mukamel1995principles} we know that the evolution of such an observable $O$ is associated with a set of susceptibilities $\chi^{O (n)}$ of different orders $n$, i.e. 
\begin{align}
    \Delta O(t)= \sum_{n=1}^{\infty}\int_{-\infty}^{\infty}\mathrm{d}s_{1} \ldots \mathrm{d}s_n &\sum_{i_{1}, ... i_{n}}\chi^{O (n)}_{i_{1} ... i_{n}}(t-s_1, ..., t - s_{n}) \nonumber\\
    &\times a_{i_{1}}(s_{1})...a_{i_{n}}(s_{n}).
    \label{eq:DeltaO}
\end{align}
Retaining a causal structure, moreover, implies the condition of $\chi^{O (n)}$ being non-zero only if $t\geq s_1 \geq...\geq s_n$. The summation indices thereby run over all perturbations to the Hamiltonian $\{X_{i}\}$. For the scenario studied in the main text of having two perturbing fields ($\vec{E}$ and $\vec{B}$) this implies $i_{1}, ...,i_{n} \in \{P, M\}$. We can therefore have two kinds of susceptibilities in the first order, four in the second order, and eight in the third order. Using the delta-like properties of the pump sequence \eqref{eq:PulseSequence} yields first and second order contributions 
\begin{align}
\Delta O^{(1)}(t)=& \chi^{O (1)}_{M}(t)B_{0} + \chi^{O (1)}_{M}
(t - \tau)B_\tau \nonumber \\
+&\chi^{O (1)}_{P}(t)E_{0} + \chi^{O (1)}_{P}(t - \tau)E_\tau
\end{align}
\begin{align}
\Delta O^{(2)}(t)=&\chi^{O (2)}_{MM}(B_0)^2  +\chi^{O (2)}_{MM}(B_\tau)^2 + \chi^{O (2)}_{MM}B_\tau B_0 \nonumber \\
+& \chi^{O (2)}_{PP}(E_0)^2 +\chi^{O (2)}_{PP}(E_\tau)^2 +\chi^{O (2)}_{PP}E_\tau E_0 \nonumber\\
+&\chi^{O (2)}_{PM}E_0 B_0 +\chi^{O (2)}_{PM}E_\tau B_\tau  +\chi^{O (2)}_{PM}E_\tau B_0 \nonumber\\
+&\chi^{O (2)}_{MP}B_0 E_0  +\chi^{O (2)}_{MP}B_\tau E_\tau  +\chi^{O (2)}_{MP}B_\tau E_0.
\label{eq:Chi2}
\end{align}
Here, we suppressed the temporal dependence in the susceptibilities in Eq.\eqref{eq:Chi2} for simpler presentation of the individual contributions. The temporal arguments can, however, be deduced from the associated field components, see \eq{eq:DeltaO}, i.e. the last contribution to \eq{eq:Chi2} with all arguments reads $\chi^{O (2)}_{MP}(t-\tau, t)B_{\tau}E_{0}$. Note that this especially implies that the first two contributions to \eq{eq:Chi2} couple to the same susceptibility $\chi^{O (2)}_{MM}$, however, evaluated at different times. Computing the linear response is achieved by setting the intensity of one pump pulse to zero, i.e. $B_{\tau}=E_{\tau}=0$ for $\Delta O_{A}$ or $B_{0}=E_{0}=0$ for $\Delta O_{B}$. The non-linear part of the response is now given by the difference 
\begin{equation}
    \Delta O^{\mathrm{NL}}(t) = \Delta O(t) - \Delta O_{\mathrm{A}}(t) - \Delta O_{\mathrm{B}}(t). 
\end{equation}
The latter only contains terms where fields evaluated at both pumping times $t=0$ and $t=\tau$ show up simultaneously. 
At second order this results in the responses for polarization and magnetization specified in \eqqs{eq:P_NL}{eq:M_NL} of the main text. At third order the polarization and magnetization responses read
\begin{widetext}
\vspace{-\baselineskip}
\begin{align}
P^{(3)}(t + \tau) &= \chi^{P (3)}_{PPP}(t,t+\tau,t+\tau)E_{\tau}(E_{0})^{2} +\chi^{P (3)}_{PPP}(t,t,t+\tau)(E_{\tau})^2 E_{0}+\chi^{P (3)}_{PPM}(t,t,t+\tau)(E_{\tau})^2 B_{0} \nonumber \\ 
&+ \chi^{P (3)}_{PPM}(t,t+\tau,t+\tau)E_{\tau}E_{0}B_{0}   + \chi^{P (3)}_{PMP}(t,t+\tau,t+\tau)E_{\tau}B_{0}E_{0} +\chi^{P (3)}_{PMP}(t,t,t+\tau)E_{\tau}B_{\tau}E_{0} \nonumber \\ 
&+\chi^{P (3)}_{MPP}(t,t+\tau,t+\tau)B_\tau(E_{0})^2  
+\chi^{P (3)}_{MPP}(t,t,t+\tau)(B_{\tau})E_{\tau}E_{0} + (M,B)\leftrightarrow (P,E).
\label{eq:P_3_NL}\\
M^{(3)}(t + \tau) &= \chi^{M (3)}_{MMM}(t,t+\tau,t+\tau)B_{\tau}(B_{0})^{2} + \chi^{M (3)}_{MMM}(t,t,t+\tau)(B_{\tau})^2 B_{0} + \chi^{M (3)}_{MMP}(t,t,t+\tau)(B_{\tau})^2 E_{0}  \nonumber \\
&+\chi^{M (3)}_{MMP}(t,t+\tau,t+\tau)B_{\tau}B_{0}E_{0} +\chi^{M (3)}_{MPM}(t,t+\tau,t+\tau)B_{\tau}E_{0}B_{0} +\chi^{M (3)}_{MPM}(t,t,t+\tau)B_{\tau}E_{\tau}B_{0}  \nonumber \\ &+\chi^{M (3)}_{PMM}(t,t+\tau,t+\tau)E_\tau(B_{0})^2 +\chi^{M (3)}_{PMM}(t,t,t+\tau)E_{\tau}B_{\tau}B_{0} + (P,E)\leftrightarrow (M,B) 
\label{eq:M_3_NL}
\end{align}
\end{widetext}
The first two contributions to \eq{eq:P_3_NL} contain only interactions with the $E$-field component of the pump and can hence be denoted as the purely electric response of the system at third order. Similarly the first two terms of \eq{eq:M_3_NL} denote purely magnetic third order response. Purely magnetic respectively electric contributions are compared to the total signal in \fig{fig3} of the main text.\\
To make quantitative predictions about the response we are left to compute the susceptibilities $\chi^{O(n)}$ from the perturbations to the Hamiltonian $\{X_{i}\}$. This can be done using a generalized Kubo's formula
\begin{align}
    \chi^{O (2)}_{X_{1}X_{2}} (t, t+\tau) =& \dfrac{i^{2}}{L} \theta(t)\theta(\tau) \nonumber \\
    & \times \langle\big[ [X_{2}(t+\tau), X_{1}(\tau)], O(0) \big]\rangle.
    \label{eq:GeneralizedKubo}
\end{align}

Generalizations to higher order susceptibilities can be obtained by enlarging the nested commutator of \eq{eq:GeneralizedKubo} to more operators.

\subsection{Selection Protocol for Response Functions}

As emphasized in the main text, the measurement response of the 2DCS protocol of \figc{fig1}{a} generically contains a superposition of different coupling contributions, including terms from correlations of one component of the electromagnetic pulses as well as cross correlations between electric and magnetic components. In the following we outline protocols that allow to differentiate between different contributions of the response. We show how to quantify the amount of polarization coupling in the system. In particular, we will highlight the consequences of (i) spatial inversion of the propagating THz pulse, and (ii) rotation of the sample on the measurement outcome.

\subsubsection*{Inverting the Propagation Direction of the THz Pulse}

The general strategy will be to invert the sign of the electric field and to superimpose the measurement outcomes of the direct and inverted fields. This can be achieved by inverting the direction of pulse propagation, i.e. $\vec{k}\to-\vec{k}$. For the field components spatial inversion is reflected in a mapping of the vector $\vec{E}\to - \vec{E}$ while for the pseudo-vector $\vec{B}\to \vec{B}$. The induced polarization $P^{\mathrm{NL}}(t\vert -\vec{E}, \vec{B})$ and magnetization $M^{\mathrm{NL}}(t\vert -\vec{E}, \vec{B})$ for the new geometry obtain a minus sign in each component, where $\vec{E}$ appears at odd orders. In case of a dimerized chain the leading order contribution to the non-linear response appears at second order and only cross correlations are affected by the changes in the experimental setup.
The individual contributions can hence be isolated by feasible superposition of results according to \eqqs{eq:P_parallel}{eq:P_cross} of the main text. 
By reversing the direction of pulse propagation, also the sign of measured electric field will flip $E^{\text{NL}}\to -E^{\text{NL}}$. Taking into account how the measured electric field adds up from induced polarization and magnetization 
\begin{equation}
    E^{\mathrm{NL}}(t\vert \vec{E}, \vec{B}) = i \bigl [ a_{E}P^{\mathrm{NL}}(t\vert \vec{E}, \vec{B}) + b_{E} M^{\mathrm{NL}}(t\vert \vec{E}, \vec{B})\bigr]
    \label{eq:E_NL}
\end{equation}
we conclude that $a_{E}, b_{E}$ transform as $a_{E}\to a_{E}$ and $b_{E}\to -b_{E}$. This results in an altered measured response 
\begin{equation}
    E^{\mathrm{NL}}(t\vert -\vec{E}, \vec{B}) = i \bigl [ a_{E}P^{\mathrm{NL}}(t\vert -\vec{E}, \vec{B}) - b_{E} M^{\mathrm{NL}}(t\vert -\vec{E}, \vec{B})\bigr]
    \label{eq:E_NL_transformed}
\end{equation}
Thus, adding and subtracting \eqq{eq:E_NL}{eq:E_NL_transformed} yields a symmetric and antisymmetric response \eqq{eq:E_NL_sym}{eq:E_NL_asym} of the main text. Explicitly writing out the symmetric component
\begin{align}
E^{\text{NL}}_{\text{sym}}(t) &= i\bigl[ a_{E} P^{\mathrm{NL}}_{\parallel} (t) +  b_{E} M^{\mathrm{NL}}_{\times} (t)\bigr] \nonumber \\
 &=  i\bigl[ a_{E} (\chi^{P (2)}_{PP}(t-\tau, t)E_{\tau}E_{0}  + \chi^{P (2)}_{MM}(t -\tau, t)B_{\tau}B_{0}) \nonumber \\
 & + b_{E} (\chi^{M (2)}_{MP}(t-\tau, t ) B_{\tau}E_{0}  + \chi^{M (2)}_{PM}(t -\tau, t)E_{\tau}B_{0}) \bigr]
 \label{eq:E_NL_sym_comp}
\end{align}
\eq{eq:E_NL_sym_comp} implies that $E^{\mathrm{NL}}_{\text{sym}}$ would vanish if the $P$ operator for the system vanishes. Thus, this protocol tells us whether the system would couple to an electric field or not.\\

\subsubsection*{Rotating the Sample}

A similar result can be obtained by rotation of the sample by $180^{\circ}$ around the axis aligned with the magnetic field. While this leaves the magnetic and electric field components unaltered, the relative orientation between pulse and sample are modulated. We denote the spatial axes of real space by $\{x_r,y_r,z_r\}$ and that of the sample by $\{x_s,y_s,z_s\}$. 
 When rotating the sample this maps $\{x_s,y_s,z_s\}\to\{-x_s,-y_s, z_s\}$. While this leaves the magnetization pointing along the rotation axis the untouched, the polarization response gets inverted. As a results the couplings transform as $E^{x},P^{x} \to -E^{x},P^{x}$ and $B^{z},M^{z} \to B^{z},M^{z}$. This yields the same response as expected from an inversion of the propagation direction of the pulse propagation discussed above.

\subsection{Jordan-Wigner Transformation}
 
To compute the 2DCS response of the TKSC we derive the diagonal version of the Hamiltonian as well as magnetization and polarization operators. This can be achieved by an exact mapping from spins to spinless fermions via Jordan-Wigner transformation. The perturbation operators commute with the parity operator, and hence, we can solve the system in the sector of even parity containing the ground state\cite{Brzezicki2007,Menchyshyn2015,Sim2023a,You2016})
 \begin{gather}
{\sigma}_{i}^{+}\equiv\dfrac{1}{2}({\sigma}_{i}^{x}+i {\sigma}_{i}^{y}) = c_{i}\exp(i\pi \sum_{j=1}^{i-1}c_{j}^{\dagger}c_{j}) \\
 {\sigma}_{i}^{z}=1-2c_{j}^{\dagger}c_{j}.
\label{eq:JW-Mapping}
\end{gather}
The Hamiltonian in the fermionic basis is given by
\begin{align}
H_{\mathrm{TK}}=-\sum_{i=1}^{L/2} \bigl[&J_1\bigl( e^{i2\theta} c_{2i-1}^\dagger c_{2i}^\dagger  +  c_{2i-1}^\dagger c_{2i} \bigr)   +  \\  &J_2\bigl(  e^{-i2\theta} c_{2i}^\dagger c_{2i+1}^\dagger  +  c_{2i}^\dagger c_{2i+1}\bigr)  + \text{H.c.} \bigr]
\end{align}
and we use the mapping to momentum space via Fourier transform,  
\begin{align}
c_{2j-1} &= \sqrt{\frac{2}{L}} \sum_{k} e^{-ikj} a_{k}, &
c_{2j} &= \sqrt{\frac{2}{L}} \sum_{k} e^{-ikj} b_{k}
\label{eq_c_c_dag_fourier}
\end{align}
with $k =2\pi n/L$ and $n=-(\frac{L}{2}-1),-(\frac{L}{2}-3) ...,(\frac{L}{2}-1)$. The discrete $k$ values have been chosen to account for anti-periodic boundary conditions in the even-parity sector. The Hamiltonian takes the form
\begin{equation}
H_{\mathrm{TK}}= \sum_{k}\bigl[ B_k a^\dagger_k b_{-k}^\dagger + A_k a^\dagger_k b_k  - A_k^* a_k b_{k}^\dagger - B_k^* a_k b_{-k}  \bigl], 
\label{eq:_Ham_in_ab}
\end{equation}
where $A_k=-(J_1+J_2e^{ik})$ and $B_k=J_2 e^{i(k-2\theta)} - J_1e^{i2\theta}$.
\begin{figure}[h]
\includegraphics[width=\columnwidth]{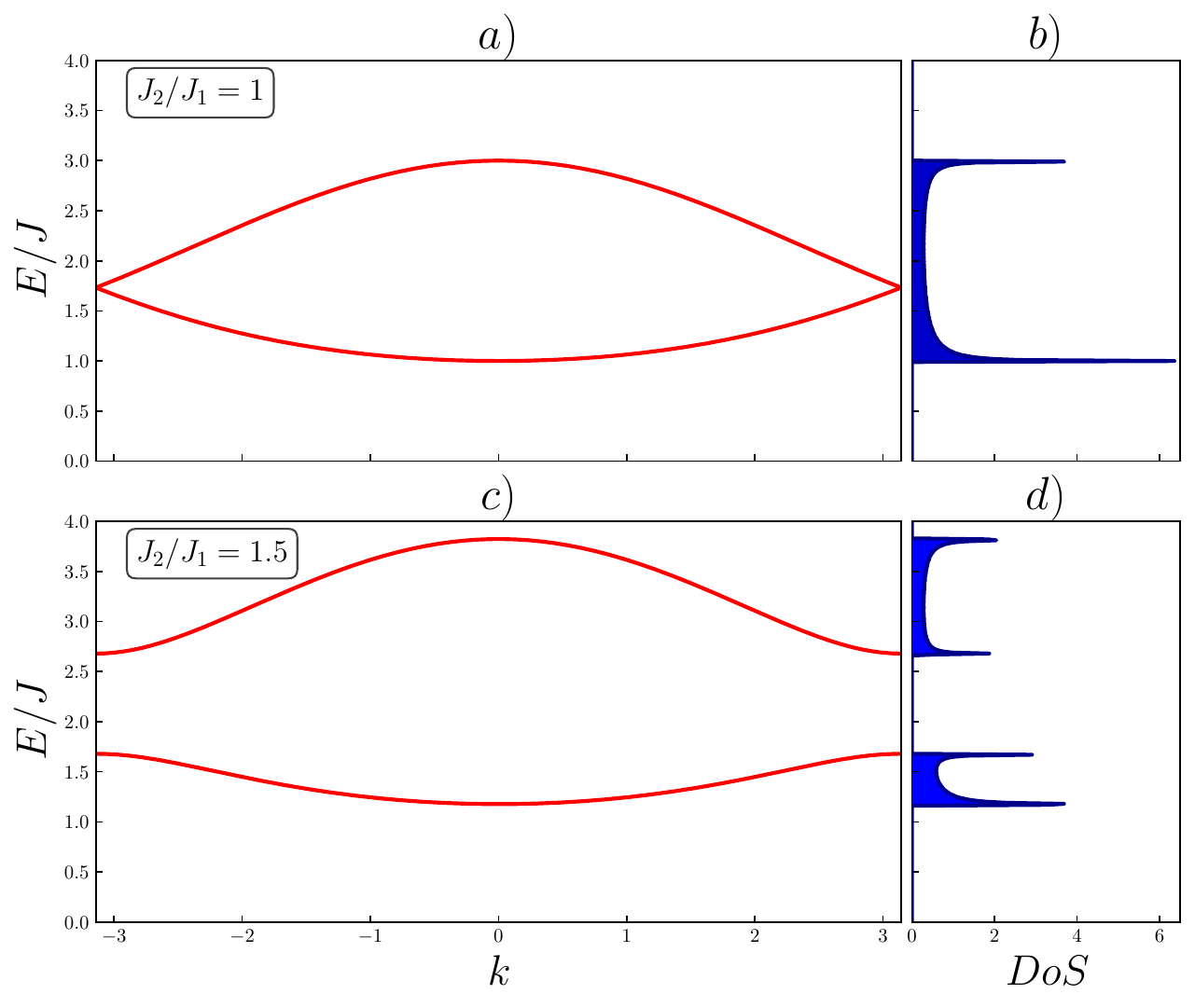}
  \caption{\textbf{Band Spectrum of $H_{\mathrm{TK}}$.} We show the dispersion (a) and density of states (DoS) (b) for an undimerized ($J_{2}=J_{1}$) TKSC with two-site unit cell. (c)-(d) Dimerization $(J_{2}= 1.5 J_{1})$ results in a gap opening between the individual bands (c) and the emergence of two additional peaks in the corresponding DoS (d).}
  \label{fig4}
\end{figure}
Writing the Hamiltonian in Bogoliubov-de-Gennes (BdG) form
\begin{gather}
H_{\mathrm{TK}} =\sum_{k>0}\psi^{\dagger}_{k}\Gamma_{k} \psi_{k} \quad \text{, with} \nonumber\\
 \Gamma_{k} = \begin{pmatrix}
0 & 0 & S_k & P_k +Q_k \\
0 & 0 & P_k - Q_k   &  - S_k\\
S_k^{\dagger} &  P_k^{\dagger} -Q_k^{\dagger}  & 0 & 0\\
 P_k^{\dagger} +Q_k^{\dagger}  &  -S_k^{\dagger}  & 0 & 0
\end{pmatrix},
\label{eq:H_spinor}
\end{gather}
where we introduced the spinor notation $\psi^{\dagger}_{k}=(a_k^{\dagger},a_{-k},b_k^{\dagger},b_{-k})$ and $S_k=-(J_1+J_2e^{ik})$, $P_k=
-i(J_1+J_2 e^{ik})\sin(2\theta)$, $Q_k=-(J_1-J_2e^{ik})\cos(2\theta)$ allows for a direct diagonalization of the form
\begin{align} 
H_{\mathrm{TK}} = \sum_{k > 0} \big[ l_k (\alpha^\dagger_k \alpha_k - \alpha_{-k} \alpha^\dagger_{-k}) + \lambda_k (\beta^\dagger_k \beta_k - \beta_{-k} \beta^\dagger_{-k}) \bigr] .
\label{eqH_diag}
\end{align} 
Here we have defined $l_k$ and $\lambda_k$ as,
\begin{align}
l_k &= \sqrt{\xi_k - \sqrt{\xi_k^2 - \tau_k^2}} &
\lambda_k &= \sqrt{\xi_k + \sqrt{\xi_k^2 - \tau_k^2}},
\end{align}
where $\xi_k = |P_k|^2 + |Q_k|^2 + |S_k|^2$ and $ \tau_k = |P_k^2 - Q_k^2 + S_k^2|$. Examples for the resulting band spectrum and the associated density of states are shown in \fig{fig4} for the case of an undimerized ($J_{2} = J_{1}$ upper panel) and dimerized  ($J_{2} = 1.5 J_{1}$ lower panel) chain.\\
Similarly, we can express the magnetization and polarization operators in the basis of free fermions. The magnetization operator is given by
\begin{align}
M^z=\dfrac{1}{2}\sum_{i=1}^{L}\sigma^{z}_{i}=\dfrac{1}{2}\sum_{i=1}^{L}(c_{i}c^{\dagger}_{i}-c^{\dagger}_{i}c_{i})
\end{align}
which can be expressed in BdG form as,
\begin{equation}
M^z=\sum_{k>0}\psi^{\dagger}_{k}\ \begin{pmatrix}
-1 & 0 & 0 & 0 \\
0 & 1 & 0  &  0\\
0 &  0  & -1 & 0\\
0  &  0 & 0 & 1
\end{pmatrix} \psi_{k}.
\end{equation}
The polarization operator through the spin-current mechanism can be expressed in a similar fashion 
\begin{align}
    P^{x}_{\mathrm{SC}}=\sum_{j}(-1)^{j}\gamma \sin(\theta)(\sigma_j^{y}\sigma_{j+1}^{x}-\sigma_j^{x}\sigma_{j+1}^{y}) \nonumber\\ 
    \propto \sum_{j=1}^{L/2} 2i(c_{2j-1}^{\dagger} c_{2j} +c_{2j-1} c_{2j}^{\dagger} - c_{2j}^{\dagger} c_{2j+1}-c_{2j} c_{2j+1}^{\dagger})
\end{align}
\begin{eqnarray}
P^x_{\mathrm{SC}} \propto \sum_{k>0} \psi_{k}^{\dagger}\begin{pmatrix}
0 & 0 & R_k & 0 \\
0 & 0 & 0  &  R_k\\
R_k^{\dagger} &  0  & 0 & 0\\
0  & R_k^{\dagger}  & 0 & 0 
\end{pmatrix}  \psi_{k} 
\end{eqnarray}
with $R_k= -2i \gamma \sin(\theta) (1+e^{ik})$.\\
The polarization operator through the exchange-striction mechanism can be found by calculating the change of the Hamiltonian due to variation of the external electric field $P=\frac{\partial H_{\mathrm{TK}}}{\partial E_{dc}}$~\cite{Brenig2023,Brenig2024}. Once we absorb the dc-electric field in rescaled exchange-interactions, we obtain
\begin{equation}
P^{x}_{\mathrm{ES}}= g\sum_{i=1}^{L/2}\bigl[ \tilde{\sigma}_{2i-1}(\theta)\tilde{\sigma}_{2i}(\theta)-\tilde{\sigma}_{2i}(-\theta)\tilde{\sigma}_{2i+1}(-\theta)\bigr]
\label{eq_PES}
\end{equation}
The corresponding spinor formulation of expression~(\ref{eq_PES}) can be obtained by setting $J_1=-g$ and $J_2=g$ in \eq{eq:H_spinor}.
\section*{Data and code availability.}
Data analysis and simulation codes are available on Zenodo upon reasonable request~\cite{zenodo}.

\bibliography{NonLinearSpectroscopy}

\section*{Acknowledgments}
We thank Wolfram Brenig, Peter Rabl, Peter Armitage and Istvan Keszmarki for encouraging discussions and especially the latter two for detailed comments on the manuscript. J.K. thanks Ribhu Kaul for helpful discussions.  We acknowledge support from the Deutsche Forschungsgemeinschaft (DFG, German Research Foundation) under Germany’s Excellence Strategy--EXC--2111--390814868, TRR 360 – 492547816 and DFG grants No. KN1254/1-2 and No. KN1254/2-1, the European Research Council (ERC) under the European Union’s Horizon 2020 research and innovation programme (grant agreement No. 851161), the European Union (grant agreement No 101169765), as well as the Munich Quantum Valley, which is supported by the Bavarian state government with funds from the Hightech Agenda Bayern Plus. 
J.K. thanks the hospitality
of Aspen Center for Physics, which is supported by National
Science Foundation grant PHY-2210452; and acknowledges support from the TUM-Imperial flagship partnership. 
A.S. acknowledges support from the Working Internship in Science and Engineering (WISE) from the Deutscher Akademischer Austauschdienst (DAAD). G.B.S was supported by Basic Science Research Program through the National Research Foundation of Korea (NRF) (RS-2024-00453943).

\section*{Author Contributions}
All authors contributed to conception, execution and write-up of this project. 

\section*{Competing Interests} 
The authors declare no competing interests.

\onecolumngrid
\newpage

\newcommand{\suppinfo}{Theory of Nonlinear Spectroscopy of Quantum Magnets}

\begin{center}
{\large Supplementary Information \\ 
\suppinfo
}\\
Anubhav Srivastava, Stefan Birnkammer, GiBaik Sim, Michael Knap, Johannes Knolle
\end{center}

\setcounter{equation}{0}  
\setcounter{figure}{0}
\setcounter{page}{1}
\setcounter{section}{0}    
\renewcommand\thesection{\arabic{section}}   
\renewcommand\thesubsection{\arabic{subsection}}   
\renewcommand{\thetable}{S\arabic{table}}
\renewcommand{\theequation}{S\arabic{equation}}
\renewcommand{\thefigure}{S\arabic{figure}}
\newcommand{\hj}{\hat{j}}
\newcommand{\hi}{\hat{i}}
\newcommand{\he}{\hat{e}}
\newcommand{\hn}{\hat{n}}

\bigskip

\section{Phase Diagram and Energy Spectrum}

\begin{figure}[h]
\includegraphics[width=9cm,height=6cm]{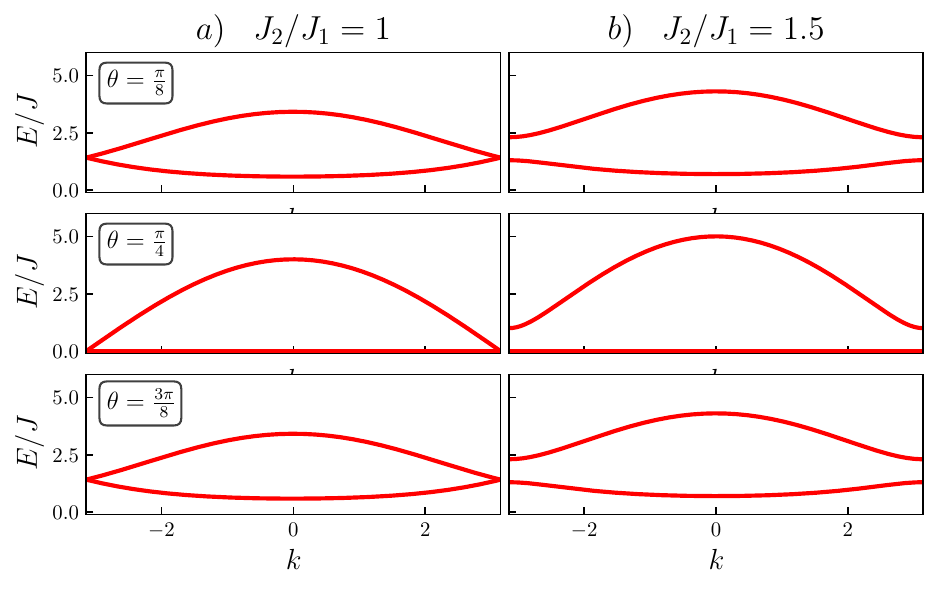}
  \caption{\textbf{Band spectra for a dimerized and undimerized TKSC}. We plot the fermionic band spectrum obtained for $H_{\mathrm{TK}}$ diagonalizing the Hamiltonian in the space of free fermions. For an (a)  undimerized ($J_{2}=J_{1}$) respectively a (b) dimerized ($J_{2}=1.5 J_{1}$) chain we show band structures for zig-zag angles $\theta=\{\frac{\pi}{8}, \frac{\pi}{4}, \frac{3\pi}{8}\}$. Both cases show similar characteristics in terms of the minimal excitation energies of the lower band, i.e. critical behavior with a flat gapless band at $\theta=\frac{\pi}{4}$. Effects of finite dimerization in (b) are, however, finite energy gaps between both fermionic bands.}
  \label{figS1:BandSpectra}
\end{figure}

\begin{figure}[!ht]
\includegraphics[width=9.1cm,height=7cm]{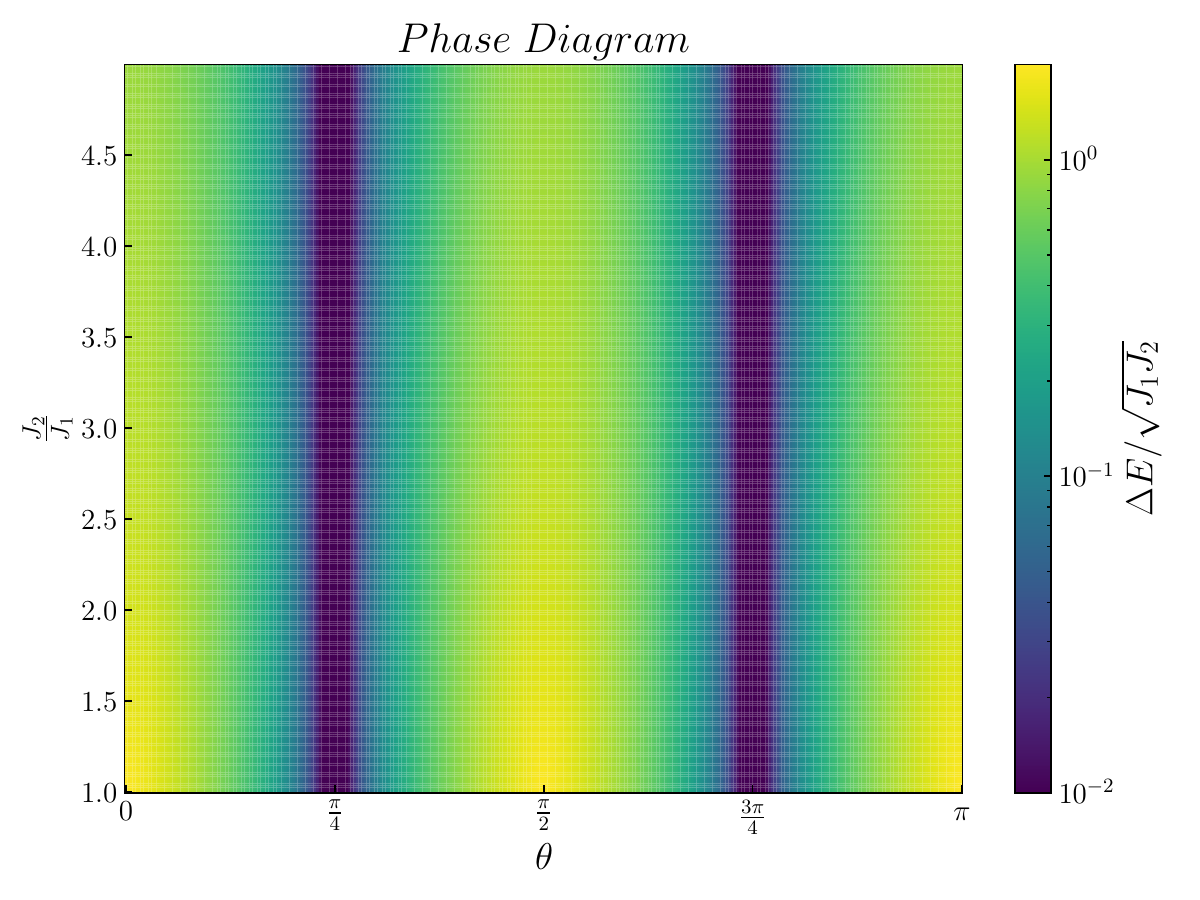}
  \caption{\textbf{
Ground state phase diagram of $H_{\mathrm{TK}}$}. We show the excitation gap $\Delta E$ in the model as a function of zig-zag angle $\theta$ and dimerization $J_{2}/J_{1}$. We identify critical lines in the phase diagram at $\theta=\frac{\pi}{4}$ and $\theta=\frac{3\pi}{4}$. Apart from that the model exhibits ferromagnetic order oriented along the $\hat{y}$-direction for $\frac{\pi}{4}<\theta <\frac{3\pi}{4}$. For all other values of $\theta$ we find ferromagnetism along the $\hat{x}$-direction. Notably, we find that changes in the dimerization do not result in gap closing and hence only smoothly transform the ground state.}
  \label{figS2}
\end{figure}

As discussed in the main text we consider the example of a twisted Kitaev spin chain (TKSC)~\cite{Menchyshyn2015, You2016}
\begin{equation}H_{\mathrm{TK}} =-\sum_{i=1}^{L/2} \bigl(J_1\tilde{\sigma}_{2i-1}(\theta)\tilde{\sigma}_{2i}(\theta)+J_2\tilde{\sigma}_{2i}(-\theta)\tilde{\sigma}_{2i+1}(-\theta)\bigr)
\label{eqHamMethods}
\end{equation}
Here $L$ denotes the number of sites in the system and $J_{1}, J_{2} > 0$ are the ferromagnetic exchange couplings. The notation $\tilde{\sigma_{i}}(\theta)=\cos(\theta)\sigma_{i}^{x}+\sin(\theta)\sigma_{i}^{y}$ indicates a rotated basis of Pauli matrices. The Hamiltonian is solved exactly by mapping spins to free fermions (see main text). The excitations in the system are hence described by dressed-domain wall or spinons. Diagonalizing the fermionic Hamiltonian with two site unit cell yields the band spectra depicted in \figcc{figS1:BandSpectra}{a}{b} for the case of an undimerized ($J_{2}=J_{1}$) respectively dimerized ($J_{2}=1.5J_{1}$) chain. We plot the band structure for different values of the zig-zag angle $\theta\in\{\frac{\pi}{8}, \frac{\pi}{4}, \frac{3\pi}{8}\}$. While we find a finite energy gap for $\theta= \frac{\pi}{8}, \frac{3\pi}{8}$, the theory at $\theta=\frac{\pi}{4}$ is critical and characterized by a flat gapless band. Interestingly, finite dimerization in the chain does not qualitatively change the characteristics of the overall energy gap, instead it only causes a gap opening between both fermionic bands at the edges of the Brillouin zone, see \figc{figS1:BandSpectra}{b}. Extracting the energy gap $\Delta E$, i.e. the minimal excitation energy for the lower fermionic band, as a function of $\theta$ and dimerization $\frac{J_{2}}{J_{1}}$ we find that the ground state phase diagram of our system contains two distinct ferromagnetic phases orientated along the $\hat{x}$- respectively $\hat{y}$-direction. By tuning the zig-zag angle $\theta$ of our chain we can tune through the different phases as emphasized in \fig{figS2}. Both ferromagnetic phases are separated by gapless lines in phase space at zigzag angles $\theta=\frac{\pi}{4}, \frac{3\pi}{4}$. As already indicated \fig{figS1:BandSpectra} we find that finite dimerization in the chain does not change the overall character of the ground state as changes in $J_{2}/J_{1}$ do not lead to gap closing. The ground state is thus given by a smooth deformation of the corresponding $\hat{x}$- or $\hat{y}$-polarized product states.

 \section{Individual contributions to the non-linear spectroscopy response}

\begin{figure}[!ht]
\centering
\hfill
\begin{minipage}[t]{0.2375\textwidth}
\includegraphics[width=\textwidth, height=6cm]{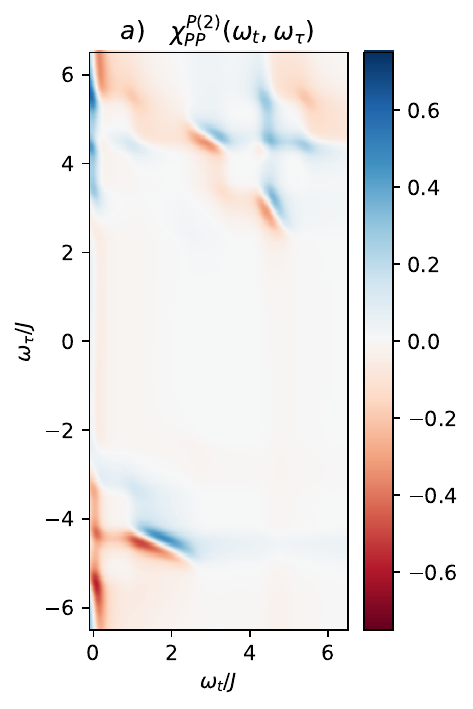} 
\end{minipage}
    \hfill
\begin{minipage}[t]{0.2375\textwidth}
\includegraphics[width=\textwidth, height=6cm]{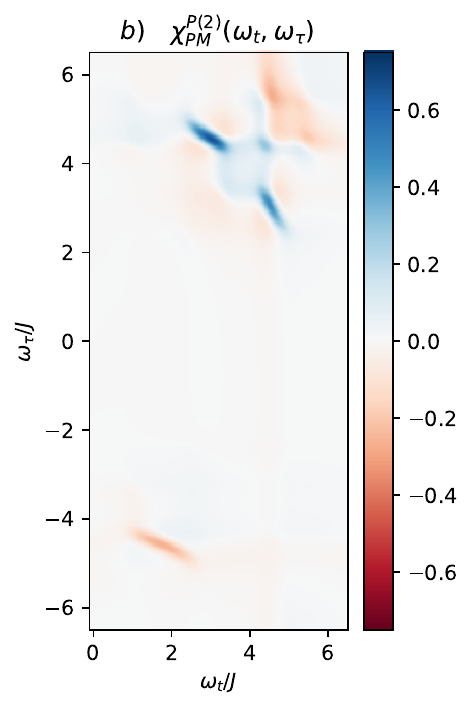} 
\end{minipage}
 \hfill
\begin{minipage}[t]{0.2375\textwidth}
\includegraphics[width=\textwidth, height=6cm] {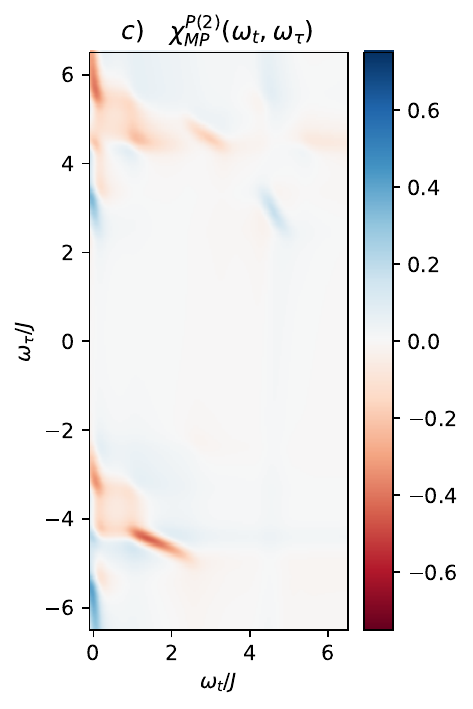} 
\end{minipage}
\hfill
\begin{minipage}[t]{0.2375\textwidth}
\includegraphics[width=\textwidth, height=6cm]{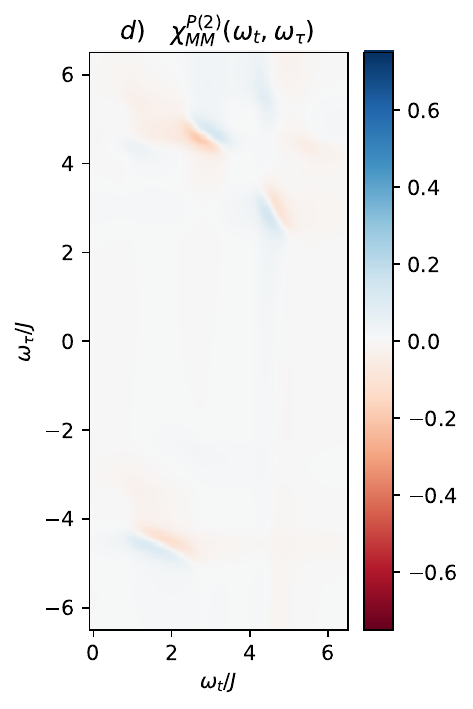} 
\end{minipage}
\caption{\textbf{Polarization response.} We show all contributions to the polarization response at second order including the purely electric response (a), cross coupling contributions (b) - (c) and the magnetic response (d).}
\label{fig3:PolarizationResponse}
\end{figure}

\begin{figure}[!ht]
\centering
\hfill
\begin{minipage}[t]{0.2375\textwidth}
\includegraphics[width=\textwidth, height=6cm]{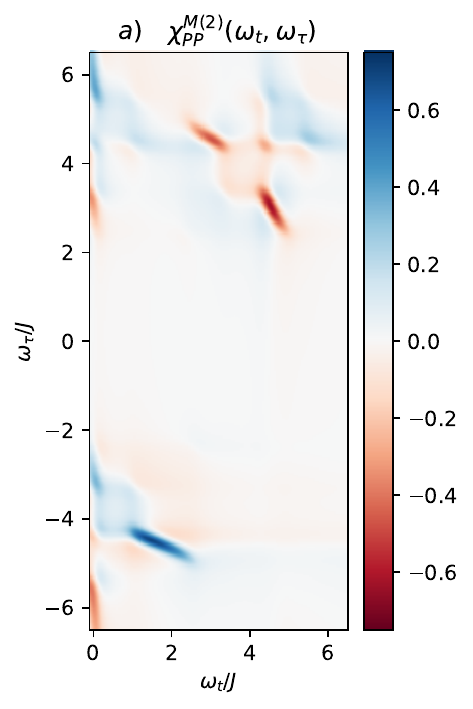} 
\end{minipage}
    \hfill
\begin{minipage}[t]{0.2375\textwidth}
\includegraphics[width=\textwidth, height=6cm]{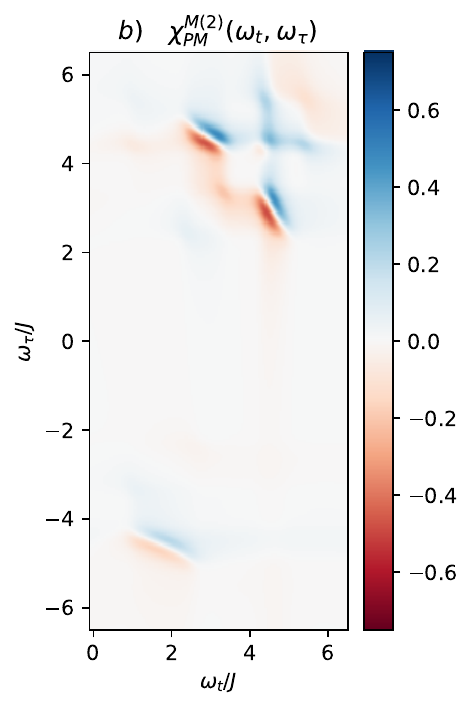} 
\end{minipage}
 \hfill
\begin{minipage}[t]{0.2375\textwidth}
\includegraphics[width=\textwidth, height=6cm] {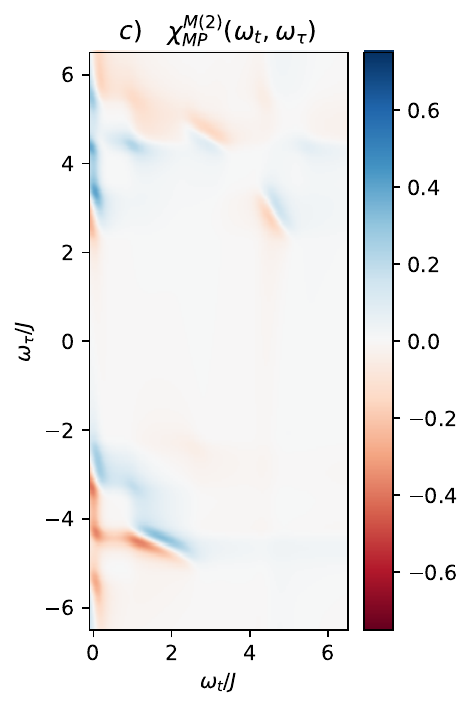} 
\end{minipage}
\hfill
\begin{minipage}[t]{0.2375\textwidth}
\includegraphics[width=\textwidth, height=6cm]{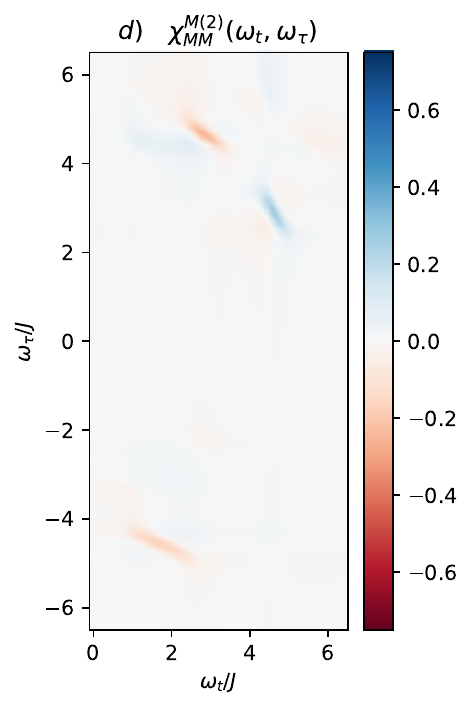} 
\end{minipage}
\caption{\textbf{Magnetization response.} We show all terms contributing to the magnetization response at second order including the purely electric response (a), cross coupling contributions (b) - (c) and the magnetic response (d).}
\label{fig4:MagnetizationResponse}
\end{figure}

\begin{figure}[!ht]

\includegraphics[width=0.3\textwidth, height=8cm]{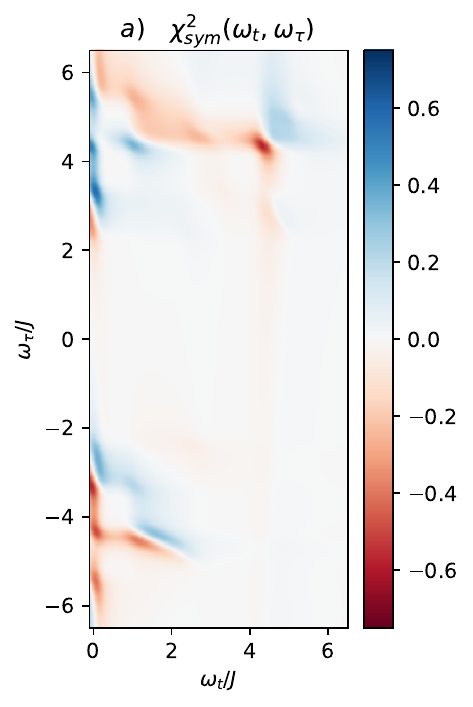} 
\includegraphics[width=0.3\textwidth, height=8cm]{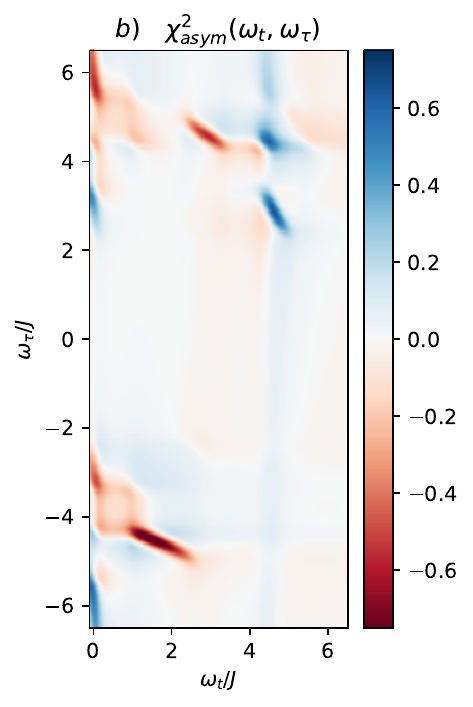} 

  \caption{ \textbf{Symmetric and anti-symmetric channel for radiated electric field.} We show the dominant second order susceptibilities $\chi^{(2)}_{\text{sym}}$ (a) and $\chi^{(2)}_{\text{asym}}$ (b) contributing to the electric field emitted by the sample. 
  }
  \label{fig5:Sym_Asym_responses}
\end{figure}

 In the main text we have emphasized that the total response measured in 2DCS experiments is added up from various contributions resulting from all different combinations of polarization and magnetization couplings. Here, we will focus on second order contributions, which usually dominate the experimental response. In general all second order terms are given by a functional form~\cite{mukamel1995principles} 
 \begin{align}
    \chi^{O (2)}_{X Y} (t, t+\tau) =& \dfrac{i^{2}}{L} \theta(t)\theta(\tau) \langle\big[ [Y(t+\tau), X(\tau)], O(0) \big]\rangle, 
    \label{eq:GeneralizedKuboMeth}
\end{align}
where all operators $O, X, Y$ represent either polarization $P^{x}$ or magnetization $M^{z}$ terms. The response spectrum is obtained from \eq{eq:GeneralizedKuboMeth} by performing a Fourier transformation in both the time delay for the second pump pulse $\tau$ as well as the evolution time $t$, see \Secc{ssec:FourierTrafo} for a precise definition. For the example of $\ch{CoNb2O6}$ with inversion symmetry breaking these operators are given by standard Zeeman coupling for magnetization $M^{z} = \sum_{i} \sigma_{i}^{z}$ and a polarization operator given by exchange striction
\begin{align}
P^{x}_{\mathrm{ES}} &\propto \sum_{i=1}^{L/2}\bigl( \tilde{\sigma}_{2i-1}(\theta)\tilde{\sigma}_{2i}(\theta)-\tilde{\sigma}_{2i}(-\theta)\tilde{\sigma}_{2i+1}(-\theta)\bigr) 
\label{eq:P_ESMethods}
\end{align}
with $\tilde{\sigma_{i}}$ as defined for the Hamiltonian~(\ref{eqHamMethods}). We show all four contributions included in the polarization response in \fig{fig3:PolarizationResponse}  and all contributions to the magnetization response in \fig{fig4:MagnetizationResponse}.
It is worth noting that while being computed individually the shown contributions are not separately measurable in experiment. Conventional 2DCS experiments will always measure the electric field emitted by the sample using electro-optical sampling. The latter is given by the sum of all terms shown in \fig{fig3:PolarizationResponse} and \fig{fig4:MagnetizationResponse} weighted by the corresponding field strength and geometric factors assigned to the different terms, see main text. Applying, however, the symmetry protocol discussed in the main manuscript allows us to, moreover, access symmetric and antisymmetric combinations given by
\begin{align}
    \chi^{(2)}_{\text{sym}}(t) &= \chi^{P (2)}_{PP}(t-\tau, t)  + \chi^{P (2)}_{MM}(t -\tau, t)
+ \chi^{M (2)}_{MP}(t-\tau, t )   + \chi^{M (2)}_{PM}(t -\tau, t)\\
\chi^{(2)}_{\text{asym}}(t) &= \chi^{P (2)}_{MP}(t-\tau, t)  + \chi^{P (2)}_{PM}(t -\tau, t) 
+ \chi^{M (2)}_{MM}(t-\tau, t )   + \chi^{M (2)}_{PP}(t -\tau, t)
\end{align}
representing the response for the case of equal geometric factors $a_{E}=b_{E}=1$ and all field strengths of the THz pulse set to $1$. Results for the response spectra $\chi^{(2)}_{\text{sym}}(\omega_{t}, \omega_{\tau})$ and $\chi^{(2)}_{\text{asym}}(\omega_{t}, \omega_{\tau})$ are shown in \figcc{fig5:Sym_Asym_responses}{a}{b}.

\section{Fourier Transform}
\label{ssec:FourierTrafo}
The data sets are obtained by evolving the system for time $t=160/J$ and $\tau=160/J$. We set $J=1$ for our simulations and calculate the response functions at time intervals of $\Delta t =\Delta \tau=0.4/J $. Therefore, we have $N^2=400^2$ points in the data set.
We then perform a discrete inverse Fourier transform to obtain Fourier components. 
For the second-order response functions, we have:
\begin{align}
    \chi^{(2)}(\omega_t,\omega_\tau)= \frac{1}{N^2}\sum_{t,\tau=0, \Delta t,..,(N-1)\Delta t} \chi^{(2)}(t,t+\tau)  e^{i  \left( \omega_t t + \omega_\tau \tau\right)}
\end{align}
As demonstrated in Eq.~(19) and (20) of the Methods section, we can have two distinct limits in the third order. The corresponding Fourier components are given by

\begin{align}
    \chi^{(3,1)}(\omega_t,\omega_\tau)= \frac{1}{N^2}\sum_{t,\tau=0, \Delta t,..,(N-1)\Delta t} \chi^{(3)}(t,t+\tau,t+\tau)  e^{i  \left( \omega_t t + \omega_\tau \tau\right)}
\end{align}
\begin{align}
    \chi^{(3,2)}(\omega_t,\omega_\tau)= \frac{1}{N^2}\sum_{t,\tau=0, \Delta t,..,(N-1)\Delta t} \chi^{(3)}(t,t,t+\tau)  e^{i  \left( \omega_t t + \omega_\tau \tau\right)}.
\end{align}
Assuming the strengths of $\delta-$peaks at $t=0$ and $t=\tau$ to be equal in the perturbing signal, see \eqqs{eq:PulseSequenceE}{eq:PulseSequence} of the main text, we can define $\chi^{(3)}(\omega_t,\omega_\tau)$ by adding the above two contributions.
\begin{align}
    \chi^{(3)}(\omega_t,\omega_\tau)= \chi^{(3,1)}(\omega_t,\omega_\tau)+\chi^{(3,2)}(\omega_t,\omega_\tau)
\end{align}

\section{Symmetry Considerations}
As we will demonstrate in the following, a lot of information about the different susceptibilities can be deduced from symmetry properties of the Hamiltonian. For the undimerized case ($J_{2}=J_{1}$) the Hamiltonian respects inversion symmetry along a bond center $\mathcal{I}$ as well as two glide symmetries $G_{x}=T_{c}\circ e^{i\pi/2\sum_{j}^{L}\sigma_{j}^{x}}=T_{c}\circ \underset{j}{\bigotimes}(i \sigma_j^x)$ and $G_{y}=T_{c}\circ e^{i\pi/2\sum_{j}^{L}\sigma_{j}^{y}}= T_{c}\circ \underset{j}{\bigotimes}(i \sigma_j^y)$. Here $T_{c}$ denotes translation by one lattice site. The two distinct ferromagnetic phases of the model, shown in \fig{figS2}, can thereby be distinguished via symmetry breaking of one of the glide symmetries. While for example the ground state of the $\hat{x}-$ferromagnetic phase transforms trivially under $\mathcal{I}$ and $G_{x}$, it breaks $G_{y}$~\cite{Sim2023a}. To analyze the response of the system it is furthermore necessary to study the transformation properties of both polarization as well as the magnetization couplings under these symmetries.

\subsection{Transformation properties of \texorpdfstring{$H_{\mathrm{TK}}$}{HTK} and \texorpdfstring{$P^{x}_{\mathrm{ES}}$}{PxES}}

To analyze the transformation properties of $H_{\mathrm{TK}}$ of~(\ref{eqHamMethods}) and $P^{x}_\mathrm{ES}$ from~(\ref{eq:P_ESMethods}) under glide symmetry $G_{x}$ it is useful to determine how $\tilde{\sigma}_{i}(\theta)=\cos(\theta){\sigma_{i}}^{x}+\sin(\theta){\sigma_{i}}^{y}$ transforms. We find $\tilde{\sigma}_{i}(\theta)\mapsto G_{x}^{-1} \tilde{\sigma}_{i}(\theta) G_{x} = \tilde{\sigma}_{i+1}(-\theta)$.
This confirms that $H_{\mathrm{TK}}$ in fact transforms trivially under $G_{x}$, while $P^{x}_\mathrm{ES}$ accumulates a minus sign from the transformation.
Conclusions for transformation with respect to $G_{y}$ follow the same strategy and yield the same symmetry properties.\\
Next, we analyze the transformation properties under inversion along a bond center $\mathcal{I}$. To this end, it is convenient to label the sites symmetrically using indices $i\in\{-\frac{L}{2}, \frac{L}{2}\}$. As a result inversion along the center bond, linking sites $0$ and $1$, takes $\vec{\sigma}_{i} \mapsto \mathcal{I}^{-1} \vec{\sigma}_{i} \mathcal{I} = \vec{\sigma}_{-i + 1}$.

\begin{align}
    \mathcal{I}^{-1}H_{\mathrm{TK}}\mathcal{I} &=-\mathcal{I}^{-1}\sum_{i=-L/4}^{L/4} \Big[ J_1\tilde{\sigma}_{2i-1}(\theta)\tilde{\sigma}_{2i}(\theta)+J_2\tilde{\sigma}_{2i}(-\theta)\tilde{\sigma}_{2i+1}(-\theta)\Big]\mathcal{I} 
    \\ \nonumber
    &=-\sum_{i=-L/4}^{L/4} \Big[ J_1\tilde{\sigma}_{-2i+2}(\theta)\tilde{\sigma}_{-2i+1}(\theta)+J_2\tilde{\sigma}_{-2i+1}(-\theta)\tilde{\sigma}_{-2i}(-\theta)\Big]
     \\ \nonumber
    &=-\sum_{j=-L/4}^{L/4} \Big[ J_1\tilde{\sigma}_{2j+1}(\theta)\tilde{\sigma}_{2j+2}(\theta)+J_2\tilde{\sigma}_{2j}(-\theta)\tilde{\sigma}_{2j+1}(-\theta)\Big]
    \\ \nonumber
    &= H_{\mathrm{TK}}
\end{align}    
Thus, $H_{\mathrm{TK}}$ is invariant under $\mathcal{I}$ for all $J_2/J_1$, and the same holds for $P^{x}_{\mathrm{ES}}$.\\

\subsection{Transformation properties of \texorpdfstring{$P^{x}_{\mathrm{SC}}$}{PxSC}}

As we will emphasize in the following the different form of
\begin{equation}
P^x_{\mathrm{SC}} =\gamma \sum_{i}^{L}(-1)^{i} \sin(\theta)(\sigma_i^{y}\sigma_{i+1}^{x}-\sigma_i^{x}\sigma_{i+1}^{y})
\end{equation}
compared to $P^{x}_{\mathrm{ES}}$ leads to different transformation properties~\cite{katsura2005spin}.
Under $G_x$, the $P^{x}_{\mathrm{SC}}$ operator transforms trivially, i.e.

\begin{align}\nonumber
G_x^{-1}P^{x}_\text{SC}G_x&=\gamma\underset{j}{\bigotimes}(-i \sigma_j^x)T_c^{-1}  \sum_{i}^{L}(-1)^{i} \sin(\theta)(\sigma_i^{y}\sigma_{i+1}^{x}-\sigma_i^{x}\sigma_{i+1}^{y})T_c \underset{j}{\bigotimes}(i \sigma_j^x) \\ \nonumber
&=\gamma\underset{j}{\bigotimes}(-i \sigma_j^x) \sum_{i}^{L}(-1)^{i} \sin(\theta)(\sigma_{i+1}^{y}\sigma_{i+2}^{x}-\sigma_{i+1}^{x}\sigma_{i+2}^{y})\underset{j}{\bigotimes}(i \sigma_j^x)
 \\ \nonumber
&=\gamma \sum_{i}^{L}(-1)^{i+1} \sin(\theta)(\sigma_{i+1}^{y}\sigma_{i+2}^{x}-\sigma_{i+1}^{x}\sigma_{i+2}^{y}) \\ \nonumber
&=P^{x}_\text{SC}.
\end{align}
Under inversion $\mathcal{I}$ the polarization $P^{x}_{\mathrm{SC}}$, however, flips sign
\begin{align}\nonumber
\mathcal{I}^{-1}P^{x}_{\mathrm{SC}}\mathcal{I}&=\gamma\mathcal{I}^{-1}\sum_{i=-L/2}^{L/2}(-1)^{i} \sin(\theta)(\sigma_i^{y}\sigma_{i+1}^{x}-\sigma_i^{x}\sigma_{i+1}^{y})\mathcal{I} \\\nonumber
&=\gamma\sum_{i=-L/2}^{L/2}(-1)^{i} \sin(\theta)(\sigma_{-i+1}^{y}\sigma_{-i}^{x}-\sigma_{-i+1}^{x}\sigma_{-i}^{y}) \\\nonumber
&=-\gamma\sum_{j=-L/2}^{L/2}(-1)^{j} \sin(\theta)(\sigma_{j}^{y}\sigma_{j+1}^{x}-\sigma_{j}^{x}\sigma_{j+1}^{y})
 \\  \nonumber
&=-P^{x}_{\mathrm{SC}}.
\end{align}
A summary of all symmetry properties of Hamiltonian $H_{\mathrm{TK}}$, polarization operators $P^{x}_{\mathrm{SC}}$ and $P^{x}_{\mathrm{ES}}$ as well as of the magnetization coupling $M^{z}$ is shown in \Tab{tab:Symmetries}.
\subsection{Even order susceptibilities for undimerized system}

As emphasized before, both the Hamiltonian $H_{\mathrm{TK}}$ and the ground state of the system $\ket{0}$ respect inversion symmetry. For both ferromagnetic phases there is, moreover, one glide symmetry which is preserved by $H_{\mathrm{TK}}$ and $\ket{0}$, while the other glide symmetry gets spontaneously broken by the ground state. Thus we can always find a combined symmetry transformation $U \equiv G_{\alpha} \circ \mathcal{I}$ which leaves $H_{\mathrm{TK}}$ and $\ket{0}$ invariant. According to \Tab{tab:Symmetries} the coupling terms $M^{z}, P^{x}_{\mathrm{ES}}$ and $P^{x}_{\mathrm{SC}}$  are, however, odd under $U$. Applying this symmetry transformation to even order susceptibilities as for example the second order contribution $\chi^{O (2)}_{XY}$ from \eq{eq:GeneralizedKuboMeth} we find

\begin{table}
\centering
\begin{tabular}{>{\centering\arraybackslash}m{0.1\textwidth} | >{\centering\arraybackslash}m{0.1\textwidth} | >{\centering\arraybackslash}m{0.1\textwidth} | >{\centering\arraybackslash}m{0.1\textwidth} | >{\centering\arraybackslash}m{0.1\textwidth}} 

 \diagbox[innerwidth=0.1\textwidth]{U}{O}& $H_{\mathrm{TK}}$ & $M^{z}$  & $P^{x}_{\mathrm{ES}}$ & $P^{x}_{\mathrm{SC}}$ 
 \\
 \hline
 $\mathcal{I}$ & +1 & +1 & +1 & -1 \\ 
\hline
$G_{x}$ & +1 & -1 & -1 & +1 \\ 
\hline
$G_{y}$ & +1 & -1 & -1 & +1 \\ 
\end{tabular}
\caption{\textbf{Symmetry Properties.} Transformation properties of undimerized Hamiltonian $H_{\mathrm{TK}}$ and coupling terms $M^{z}, P^{x}_{\mathrm{ES}}$ and $P^{x}_{\mathrm{SC}}$ under the symmetries $\mathcal{I}, G_{x}, G_{y}$. Entries denote the eigenvalues $\xi$ under transformation $O \mapsto U^{-1} O U = \xi O$.}
\label{tab:Symmetries}
\end{table} 
\begin{align} \nonumber
    \chi^{O (2)}_{XY}(t, t+\tau) &=\dfrac{i^{2}}{L} \theta(t)\theta(\tau)  \times \langle 0|U \big[ [U^{-1}Y(t+\tau)U, U^{-1}X(\tau)U], U^{-1}O(0)U \big] U^{-1}|0\rangle \\ \nonumber
    &=\dfrac{i^{2}}{L} \theta(t)\theta(\tau)  \times \langle 0|\big[ [-Y(t+\tau), -X(\tau)], -O(0) \big]|0\rangle \\ \nonumber  
    &=-\chi^{O (2)}_{XY} (t, t+\tau),  
\end{align}
where we made use of the fact that $U$ commutes with the time evolution of the operators $X, Y, O \in \{M^{z}, P^{x}_{\mathrm{ES}}, P^{x}_{\mathrm{SC}}\}$ in the Heisenberg picture governed by $H_{\mathrm{TK}}$ and applied the transformation properties of \Tab{tab:Symmetries}. As a result all even order susceptibilities are expected to vanish as long as $H_{\mathrm{TK}}$ respects glide symmetries as well as inversion. As discussed in the main text we can explicitly break the glide symmetry of the model by considering finite dimerization $(J_{2}\neq J_{1})$ which will result in finite contributions to the non-linear response at second order.

\subsection{Second order susceptibilities for dimerized system}
If we take $P=P_{\mathrm{SC}}$ even for the dimerized system, all second-order susceptibilities with a polarization index vanish. That $\chi^{P(2)}_{PP}$, $\chi^{M(2)}_{MP}$ , $\chi^{M(2)}_{PM}$ and $\chi^{P(2)}_{MM}$ must vanish, follows from symmetry arguments. As shown before, $P_{\mathrm{SC}} $ is odd under $\mathcal{I}$, while $M^z$ and ground state transform trivially under $\mathcal{I}$. Thus, any susceptibility with $P_{\mathrm{SC}}$ appearing an odd number of times vanishes. This can be demonstrated explicitly
\begin{align} \nonumber
    \chi^{P (2)}_{MM}(t, t+\tau)  &= \dfrac{i^{2}}{L} \theta(t)\theta(\tau) \nonumber 
    \times \langle\big[ [M(t+\tau), M(\tau)], P(0) \big]\rangle\\\nonumber &=\dfrac{i^{2}}{L} \theta(t)\theta(\tau)  \times \langle 0|\mathcal{I} \big[ [\mathcal{I}^{-1}M(t+\tau)\mathcal{I}, \mathcal{I}^{-1}M(\tau)\mathcal{I}], \mathcal{I}^{-1}P(0)\mathcal{I} \big] \mathcal{I}^{-1}|0\rangle \\ \nonumber
    &=\dfrac{i^{2}}{L} \theta(t)\theta(\tau)  \times \langle 0|\big[ [M(t+\tau), M(\tau)], -P(0) \big]|0\rangle \\ \nonumber  
    &=-\chi^{P (2)}_{MM} (t, t+\tau),  
\end{align}
Similarly, we can show $\chi^{P(2)}_{PP}$, $\chi^{M(2)}_{MP}$ and $\chi^{M(2)}_{PM}$ vanish.\\
It remains to show $\chi^{M(2)}_{PP}$, $\chi^{P(2)}_{MP}$ and $\chi^{P(2)}_{PM}$ vanish. As checked by analytic calculation, when expressed in the eigenbasis of the Hamiltonian, time evolved $P_{\mathrm{SC}}$ is anti-diagonal, and $M^z$ has vanishing diagonal terms. When such matrices are substituted in the nested commutator formulae for $\chi^{M(2)}_{PP}$, $\chi^{P(2)}_{MP}$ and $\chi^{P(2)}_{PM}$ , the resulting matrix has a vanishing diagonal. Taking the expectation value over the ground state involves summing over those diagonal entries that correspond to negative energies. Since all the diagonal entries are zero, this implies that the response functions vanish.\\
This is not the case for $P=P_{\mathrm{ES}}$. Hence, we used  $P=P_{\mathrm{ES}}$ for second-order calculations in the main text.

\end{document}